\pdfoutput=1
\documentclass[a4paper, 11pt]{article}
\usepackage[english]{babel}
\usepackage[a4paper, left=2.0cm, right=2.0cm, top=2.0cm, bottom=2.5cm,]{geometry}

\usepackage{lmodern}
\usepackage[T1]{fontenc}

\usepackage{amsmath}
\usepackage{amssymb}
\usepackage{pifont}
\usepackage{mathabx}
\usepackage{graphics}
\usepackage{graphicx}
\usepackage{tabularx}
\usepackage{bm}
\usepackage{color}
\usepackage{units}
\usepackage{blindtext}
\usepackage{textcomp}
\usepackage{float}
\usepackage{xcolor}
\usepackage{units}
\usepackage{datetime}
\usepackage{booktabs}
\usepackage{array}
\usepackage{pdflscape}	
\usepackage{titlesec}
\usepackage{multirow}
\usepackage{multicol}
\usepackage{subcaption}
\usepackage{tikz}
\usetikzlibrary{shapes,shapes.geometric,arrows}

\usepackage{fancyhdr}
\usepackage{setspace}
\usepackage{lineno}	
\usepackage{lastpage}
\usepackage{amsthm}
\theoremstyle{plain}

\newtheorem{definition}{Definition}
\newtheorem{problem}{Problem}
\usepackage[colorlinks,linkcolor=blue,citecolor=blue,urlcolor=blue]{hyperref} 
\usepackage{cleveref}

\titleformat{\section}{\normalsize\bfseries\scshape}{\thesection.}{.5em}{}
\titlespacing\section{0pt}{1ex plus .5ex minus .5ex}{-1ex plus 0.2ex}

\titleformat{\subsection}{\normalsize\bfseries\scshape}{\thesubsection.}{.5em}{}
\titlespacing\subsection{0pt}{1ex plus .5ex minus .5ex}{-1ex plus .2ex}

\titleformat{\subsubsection}{\normalsize\bfseries\scshape}{\thesubsubsection.}{.5em}{}
\titlespacing\subsubsection{0pt}{1ex plus .5ex minus .5ex}{-1ex plus .2ex}

\makeatletter
\g@addto@macro\normalsize{
	\setlength{\abovedisplayskip}{5pt}
	\setlength{\belowdisplayskip}{5pt}
	\setlength{\abovedisplayshortskip}{5pt} 
	\setlength{\belowdisplayshortskip}{5pt} 
}
\makeatother

\addto\captionsenglish{}

\crefname{figure}{Fig.}{Figs.}
\crefname{table}{Table}{Tables}
\crefname{equation}{Eq.}{Eqs.}
\Crefname{equation}{Equation}{Equations}
\crefname{problem}{Problem}{Problems}
\crefname{chapter}{Chapter}{Chapters}
\crefname{appendix}{Appendix}{Appendices}
\crefname{section}{Section}{Sections}
\crefname{subsection}{Section}{Sections}
\crefname{subsubsection}{Section}{Sections}

\newcounter{savefootnote}
\newcounter{symfootnote}
\newcommand{\symfootnote}[1]{%
	\setcounter{savefootnote}{\value{footnote}}%
	\setcounter{footnote}{\value{symfootnote}}%
	\ifnum\value{footnote}>8\setcounter{footnote}{0}\fi%
	\let\oldthefootnote=\thefootnote%
	\renewcommand{\thefootnote}{\fnsymbol{footnote}}%
	\footnote{#1}%
	\let\thefootnote=\oldthefootnote%
	\setcounter{symfootnote}{\value{footnote}}%
	\setcounter{footnote}{\value{savefootnote}}%
}

\renewenvironment{abstract}{\vspace{2ex}\hrule
	\begin{small}
		\textbf{Abstract}.}{\end{small}
	\vspace{3ex}
	\hrule}
\newenvironment{keywords}{\noindent\begin{footnotesize}\textbf{Keywords:}}{\end{footnotesize}}

\renewcommand{\vec}[1]{\bm{#1}}
\newcommand{\mat}[1]{\mathbf{#1}}
\newcommand{\set}[1]{\mathcal{#1}}
\newcommand{\R}{\mathbb R}
\newcommand{\defn}{\triangleq}
\newcommand{\transp}{^\mathrm{T}}
\newcommand{\norm}[1]{\left\lVert#1\right\rVert}
\newcommand{\yes}{\checkmark}
\newcommand{\no}{\ding{55}}

\newcommand{\ind}[1]{_{\mathrm{#1}}}
\newcommand{\hl}[1]{``#1''}
\newcommand{\dashedline}[1][2cm]{\tikz[baseline] \draw[densely dashed, red, line width=0.5pt] (0,0.5ex) -- (1.8ex,0.5ex);}

\renewcommand{\hl}[1]{#1}
\newcommand{\eg}{\textit{e.g.}}
\newcommand{\ie}{\textit{i.e.}}
\newcommand{\cf}{\textit{cf.}}
\newcommand{\abk}[1]{(#1)}

\begin{document}

\phantom{x}\vspace{-11ex}
\begin{center}
	\begin{Large}
		\textbf{
			Nonlinear Model Order Reduction of Dynamical Systems in Process Engineering: 
			Review and Comparison
		}
	\end{Large} 
	
	Jan C.~Schulze$^{a}$
	and 
	Alexander Mitsos$^{b,a,c,}$\symfootnote{Corresponding author: \tt{amitsos@alum.mit.edu}}
\end{center}

\begin{scriptsize}
	\begin{center}
		$^a$ Process Systems Engineering (AVT.SVT), RWTH Aachen University, 52074 Aachen, Germany \\
		$^b$ JARA-CSD, 52056 Aachen, Germany \\
		$^c$ Energy Systems Engineering (ICE-1), Forschungszentrum J\"ulich, 52425 J\"ulich, Germany \\
	\end{center}
\end{scriptsize}

\singlespacing

\begin{abstract}
Computationally cheap yet accurate dynamical models are {a key requirement} for real-time capable nonlinear optimization and model-based control.
When given a computationally expensive high-order prediction model,
a reduction to a lower-order simplified model can enable such real-time applications.
Herein, we review nonlinear model order reduction methods and provide a comparison of method {characteristics}.
Additionally, we discuss both general-purpose methods and tailored approaches for chemical process systems and we identify similarities and differences between these methods.
As {machine learning} manifold-Galerkin approaches currently do not account for inputs in the construction of the reduced state subspace, we extend these methods to dynamical systems with inputs.
In a comparative case study, we apply eight established model order reduction methods to an air separation process model: POD-Galerkin, nonlinear-POD-Galerkin, manifold-Galerkin, dynamic mode decomposition, Koopman theory, manifold learning with latent predictor, compartment modeling, and model aggregation.
Herein, we do not investigate hyperreduction{, \ie,} {reduction of floating point operations}.
Based on our findings, we discuss strengths and weaknesses of the model order reduction methods.
\end{abstract}

%\keywords{model simplification \and hybrid model \and invariant manifold \and nonlinear Galerkin \and non-autonomous system}
\iftrue
\begin{keywords}
model simplification, hybrid model, invariant manifold, nonlinear Galerkin, non-autonomous system
\end{keywords}
\fi

\section{Introduction}
\label{sec:intro}
Process systems in chemical engineering are inherently nonlinear, {\eg}, as a result of reaction kinetics, transport phenomena, and thermodynamic relationships \cite{Luyben.2007}.
Over the past decades, detailed mechanistic process modeling has become increasingly common in (chemical) process systems engineering.
In particular, mechanistic first-principles models, accounting for conservation of mass, energy, and (less often) momentum, along with constitutive relations, possess favorable generalization capabilities.
Hence, \hl{digital twins} and \hl{digital shadows} have been established as a valuable tool for process monitoring and operator training \cite{Pantelides.2013,Kender.2021}.
However, in such predictive models,
the high number of nonlinear differential equations is usually computationally prohibitive for online optimization and model-based control methods such as nonlinear model predictive control.
Furthermore, process models are often stiff, making them even more challenging to handle \cite{Brenan.1995}.

\begin{table}[t!]
    \centering
    \caption{{List of acronyms used in this work.}}
    \label{tab:abbrev}
    {
    \begin{normalsize}
    \begin{tabular}{ll}
    \toprule
    \textbf{Acronym} & \textbf{Term} \\
    \midrule
    ANN & Artificial neural network \\ 
    AEN & Autoencoder network \\  
    AGG & Aggregation method \\ 
    COMP & Compartment method \\ 
    DAE & Differential-algebraic system of equations \\ 
    DMD & Dynamic mode decomposition\\ 
    DMDc & DMD with controls (or general inputs)\\ 
    FOM & Full-order model\\ 
    HX & Heat exchanger\\ 
    KW & Koopman-Wiener model\\ 
    MFL & Manifold learning\\ 
    MFL-ANN & MFL combined with latent ANN predictor\\ 
    MFLu & Manifold learning with inputs \\ 
    MOR & Model order reduction \\ 
    ODE & Ordinary differential system of equations \\ 
    POD & Proper orthogonal decomposition \\ 
    POD-Res & POD residualization \\ 
    ROM & Reduced order model \\ 
    RMSE & Root mean squared error \\ 
    SPT & Singular perturbation theory \\ 
    SVD & Singular value decomposition \\ 
    \bottomrule
    \end{tabular}
    \end{normalsize}
    }
\end{table}

Model reduction is a popular technique to reduce the computational cost of optimization by means of model simplification.
In this context, a reduction in the number of differential states is called \hl{model order reduction} \abk{MOR} and the resulting model is a \hl{reduced-order model} \abk{ROM} \cite{Marquardt.2002,Antoulas.2005b}.
Conversely, the original model is the \hl{full-order model} \abk{FOM}.
MOR is particularly used for spatially distributed systems, which are commonly described by partial differential equations or modeled as networks of lumped systems. 
Examples of high-order {process} systems include computational fluid dynamics \cite{Rowley.2017},
distillation columns \cite{Wong.1980}, 
heat exchangers \cite{Haider.2020}, 
chemical reactors \cite{Georgakis.1977}, 
and integrated process networks \cite{Daoutidis.2015}.
Fine-grained mechanistic modeling of such systems can result in thousands or millions of states.

Although mechanistic models can be very high-dimensional, FOMs often exhibit a certain coherence characterized by a few dominant mechanisms or patterns, {\eg}, due to stiffness of the system \cite{Okeke.2015}.
Provided that a part of the transients vanish after a short period of time due to dissipative phenomena with a comparably fast exponential decay, we obtain a time-scale separation between fast stable decay and slow response \cite{Graham.1996}.
As a result, the high-dimensional state evolves most of the time near a low-dimensional topological subspace, 
referred to as \hl{slow manifold} \cite{Nijmeijer.1990}.
{Because this subspace is characterized by a set of invariance equations, \ie, the (quasi-)steady-state conditions of the fast stable phenomena, the slow manifold is a special type of \hl{invariant manifold} \cite{Fenichel.1971,Roberts.1989b,Krauskopf.2005}.}
An accurate ROM is obtained when finding a low-order approximation of the FOM on the slow manifold.
Such reducible systems are also called \hl{lumpable} \cite{Horstmeyer.2016}. 

MOR methods construct low-order approximations of a FOM, {\ie}, a ROM, either by projecting the FOM onto the slow manifold \cite{Shvartsman.1998,Baur.2014} or by identifying an empirical ROM directly from state data \cite{Schmid.2010,Kramer.2024}.
Projecting the FOM onto a linear subspace is called a \hl{Galerkin projection} \cite{Antoulas.2005}, a concept of high significance for this article.
{Below, we also discuss nonlinear variants of Galerkin methods as well as other projection methods.}
In many cases, a considerable reduction (removing $\unit[80]{\%}$ of the states or more) of a FOM is possible at a minor sacrifice of accuracy.
However, the range of validity of a ROM is considerably reduced as well, {\eg}, highly transient situations, such as process startup or shutdown, are {often} no longer covered.

Reduction methods can be classified according to how the FOM is utilized \cite{Rozza.2022}.
Specifically, intrusive methods rely on structural and parametric knowledge of the FOM.
{Note that the term ``intrusive'' as adopted here is fully independent from its use in structural mechanics \cite{Mignolet.2013}, and as such partially inconsistent.}
A large subclass of intrusive methods are reduced-basis methods, which seek a low-dimensional set of basis vectors spanning a linear or affine subspace on which the system approximately evolves, followed by an (intrusive) projection of the FOM onto this subspace \cite{Rathinam.2003,Ohlberger.2015}.
Non-projection-based intrusive methods include simplifying assumptions and heuristic reduction.
On the other hand, \hl{non-intrusive methods} are fully detached from the FOM equations, {\ie}, model-free.
These methods only require time series data {of the model states} to generate a ROM.

Many projection-based approaches follow a two-stage procedure to derive a ROM. 
First, the nonlinear system is reversibly transformed into a favorable representation, {\eg}, some canonical or normal form \cite{Moore.1981,Shaw.1991,Roberts.2008}.
Subsequently, reduction is performed on the transformed system by systematic truncation or residualization of dynamic components. 
Depending on the original model and the reduction method, the resulting ROM can be anything from a white-box (all equations have a physical meaning) to an almost black-box (purely empirical) model.
Examples of the first step include proper orthogonal decomposition \cite{Berkooz.1993},
{nonlinear normal modes \cite{Shaw.1991,Vizzaccaro.2021},}
input-output balancing \cite{Moore.1981,Hahn.2002},
singular perturbation canonical form \cite{Khalil.2002},
Weierstrass canonical form \cite{Romijn.Diss}, 
canonical bilinear or polynomial form \cite{Phillips.2003,Feng.2004,Gu.2011},
and Koopman lifting \cite{Mezic.2005,Williams.2015}. 

\begin{figure}[t!]
	\centering
	\includegraphics[width=0.5\linewidth]{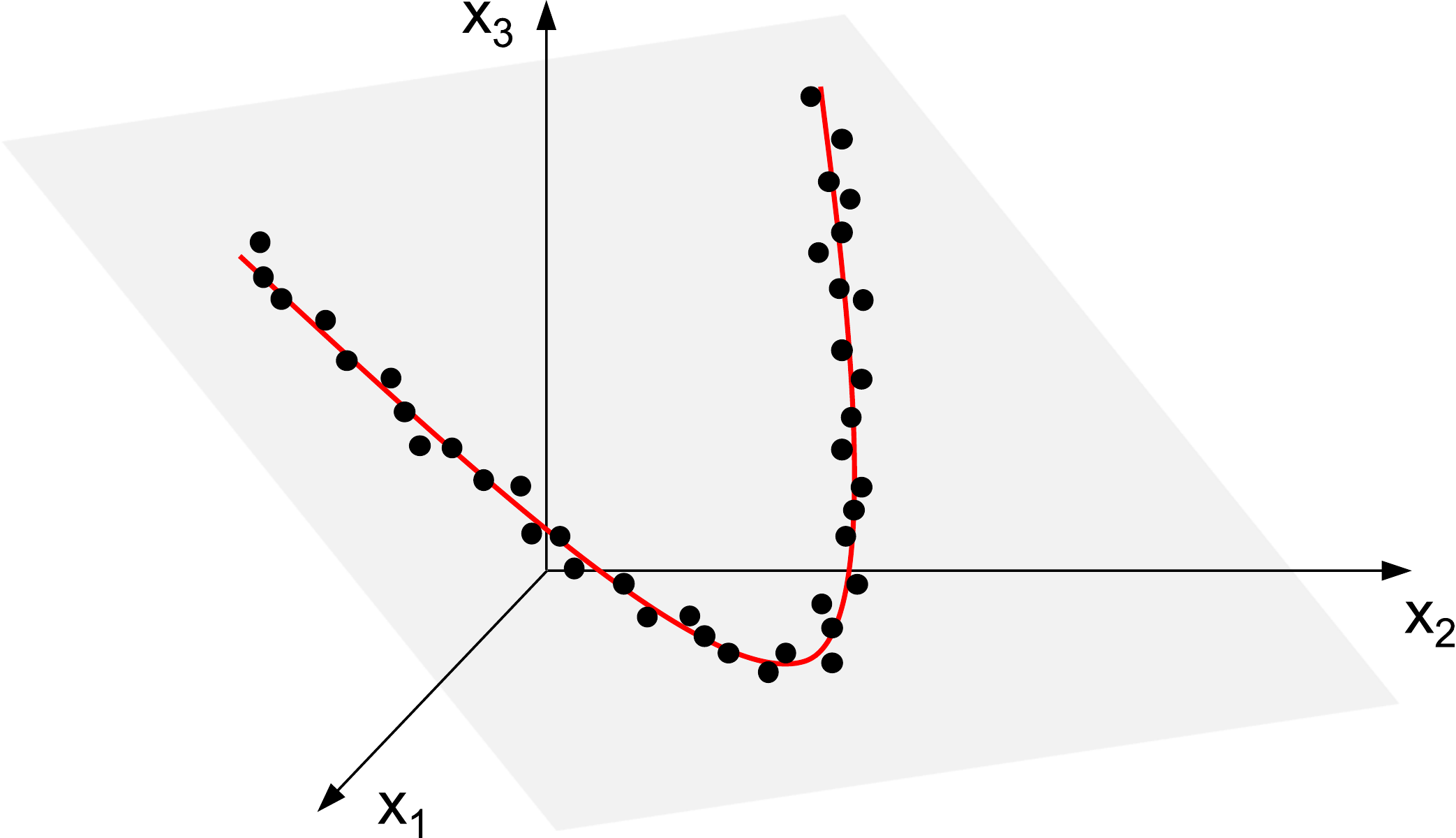}
	\caption{{Illustration of l}inear ({$\set M_1$,} $\textcolor{lightgray}{\blacksquare}$) {and} nonlinear ({$\set M_2$, }\dashedline{}) {embedded} submanifolds {of $\R^3$}.
    {The elements of $\set M_2$ are close to $\set M_1$ but not necessarily elements thereof.
    Here, the samples from the original system}
    ($\bullet$)
    {are captured similarly well by both $\set M_1$ and $\set M_2$ despite their different dimensions.}
    }
	\label{fig:subspace}
\end{figure}

For some types of systems, {\eg}, convection-dominated systems {and} wave-type equations, the assumption of a low-dimensional linear subspace does not hold \cite{Barnett.2022,Hesthaven.2022}.
Such systems live on a nonlinear submanifold rather than on a linear subspace \cite{Jauberteau.1990,Shaw.1991}.
Consequently, Galerkin projection requires a large number of basis vectors to retain high ROM accuracy \cite{Shvartsman.1998,Lee.2020}.
Instead, nonlinear subspace methods can approximate the nonlinear slow manifold more closely.
Projecting the FOM onto the nonlinear submanifold 
is a \hl{nonlinear Galerkin} \cite{Jauberteau.1990} or \hl{manifold-Galerkin} \cite{Lee.2020} projection{, which is closely connected to the concept of nonlinear normal modes in mechanics \cite{Shaw.1991}.}
Figure \ref{fig:subspace} illustrates the advantage of {using} a nonlinear manifold {as the reduced subspace}.
{Although the dimensionality of the 1D nonlinear subspace is lower than the 2D linear subspace,
the data points generated by the 3D full-order system are well captured by either submanifold.
Consequently, using a nonlinear subspace method may enable a similarly accurate ROM of lower order.}

When reducing a nonlinear FOM, the CPU cost of evaluating a ROM is often not significantly decreased, despite the lower differential order \cite{Rathinam.2003}.
Thus, to achieve a reduction in computational costs, an additional consecutive \hl{hyperreduction} step is performed \cite{Hahn.2002b,Ryckelynck.2009}, which reduces the arithmetic complexity{, \ie, floating point operations per model call}.
Hyperreduction is achieved by applying supervised approximation techniques to replace expensive terms by, {\eg}, piecewise linear functions \cite{Barrault.2004,Rewienski.2006},
polynomials \cite{Bos.2004,Nguyen.2008,Chaturantabut.2010},
sparse regression models \cite{SINDy},
or artificial neural networks \abk{ANNs} \cite{Shvartsman.2000,Hahn.2002b}.
In some cases, semi-empirical shortcut methods are an alternative to eliminate steady-state equations of a ROM \cite{Diwekar.1994}.
Further, hyperreduction methods may also be implemented in an adaptive fashion \cite{Hedengren.2005,Luthje.2021}.
In case of a differential-algebraic ROM,
hyperreduction automatically eliminates algebraic equations, so a prior reduction of algebraic variables \cite{Lee.1991,Sun.2005} is often not required.
Combining data-driven and mechanistic elements, the resulting ROM can be considered a \hl{hybrid model}.
Hyperreduction drives the ROM further towards a black-box model but is essential for real-time capability. 
The overall model reduction process thus typically involves the two ingredients (i) MOR and (ii) hyperreduction.

Although non-intrusive reduction methods do not operate on the FOM equations,
these methods are still based on {FOM} state information (and sometimes also on time derivative data).
Conversely, the identification of a {low-order} black-box model from input-output data {(\eg, using subspace identification \cite{Qin.2006}, Loewener framework \cite{Mayo.2007,Peherstorfer.2017},
or AAA algorithm \cite{Nakatsukasa.2018,Rodriguez.2023,Bradde.2025})}
does not involve {the original FOM} state{s} and thus lacks the {full} information carried by the {FOM} states.
In particular, {FOM} state information turn{s} out to be highly valuable {in some} {MOR problems}, where input-output system identification may {fail} to extract sensible latent information \cite{Schulze.2022a}. 

Despite a long history of research, 
MOR remains an evolving field with major
contributions made in recent years, {\eg}, \cite{Schmid.2010,Astolfi.2010,Williams.2015,Lee.2020}.
Comprehensive reviews of MOR are found in \cite{Givon.2004,Baur.2014,Rowley.2017,Kumar.2023c,Mallick.2025} 
as well as textbooks on linear \cite{Obinata.2001,Antoulas.2005} and nonlinear MOR \cite{Noack.2011,Holmes.2012,Benner.2017,Touze.2025}. 
However, {although} an extensive body of literature {is available},
low-barrier introductory texts \cite{Marquardt.2002,Antoulas.2005b} and comparative studies in the {process systems} field \cite{Schlegel.2002,Hahn.2002c,Rowley.2017} lack a discussion of some recent developments.
Furthermore, reduced modeling strategies that have been developed in the chemical engineering community are often not discussed in the general model reduction literature.

Over the years, managing the growing number of MOR options and making a suitable choice has become increasingly difficult.
Therefore, this article aims to provide an overview of classical and more recent methods.
We consider systems with inputs and focus on nonlinear state-space models, but we also include index-one differential-algebraic systems. 
We compare common methods both theoretically and in {an industrially relevant} case study.

As we target ROM application within {numerical optimization and model predictive control}, we want to preserve the knowledge of all states and retain flexibility in formulating the optimal control problem.
Hence, we aim for ROMs that allow reconstruction of all original states and thus focus on input-to-state reduction methods rather than input-output methods.
Consequently, we do not review input-output balancing methods for linear systems \cite{Moore.1981,Pernebo.1982}, empirical {balancing of} nonlinear systems \cite{Scherpen.1993,Lall.1999,Hahn.2002,Himpe.2014}, 
or data-driven moment matching \cite{Scarciotti.2017}.
{However, we remark that for problems with fixed input-output configuration, balancing methods can provide very low-order ROMs of high accuracy and guaranteed stability preservation \cite{Kawano.2026}.}
Further, we disregard MOR methods based on linearized FOMs, {\eg}, 
applying feedback linearization before reduction \cite{Baader.2023}.
{Finally, we remark that model reduction of non-autonomous systems with inputs shares some conceptual similarities with MOR of stochastic dynamical systems \cite{Roberts.2008,Dsilva.2016}.
Specifically, both systems feature a-priori unknown external variables.
However, we do not investigate these connections further.
}

The manuscript is structured as follows.
First, we {present} a general version of the MOR problem for systems with inputs in Section \ref{sec:mor-problem}.
{Therein, we characterize the reduced subspace based on the invariance equations and explicitly account for system inputs.}
In Section \ref{sec:methods}, we provide an overview of general-purpose MOR methods.
We divide this review into intrusive methods in Section \ref{sec:intrusive} and non-intrusive methods in Section \ref{sec:non-intrusive}.
During the review of intrusive methods{, in Section \ref{sec:manifold-learning},} we observe that current manifold learning methods do not account for external inputs.
Hence, we propose an extension {of the machine learning manifold-Galerkin method to non-autonomous systems with inputs} in Section \ref{sec:ml-control}.
The review of general-purpose methods concludes with a theoretical comparison in Section \ref{sec:comparison}.
In Section \ref{sec:pse}, we review problem-specific MOR approaches that have been developed in the field of {process systems engineering} and draw connections to methods in Section \ref{sec:methods}.
Additionally, we review applications of MOR for three important unit operations, namely distillation columns, heat exchangers, and reactors.
Section \ref{sec:case-study} provides a comparative case study, where we reduce the process model of an air separation unit.
Therein, we compare eight MOR approaches that are selected as a representative ensemble of the methods reviewed.
Notably, we limit ourselves to MOR and do not consider hyperreduction and CPU costs of integration to avoid an extra level of methods and complexity. 
However, we expect that any ROM of sufficiently low order can be made real-time capable by appropriate hyperreduction.
In Section \ref{sec:case-study:comparison}, we summarize and discuss the strengths and weaknesses of the MOR methods.
Section \ref{sec:conclusions} concludes the manuscript.

\paragraph{Notation.}
Given a matrix $\mat A \in \R^{n\times m}$, $\mat A = [\vec a_1, ..., \vec a_m]$, the linear space spanned by the column vectors $\vec a_i \in \R^n$ is denoted by $\set A = \mathrm{col}(\mat A)$ and called the \hl{column space} of $\mat A$.
{For brevity, we prefer this notation over $\set A=\mathrm{span}\{\vec a_1, ..., \vec a_m\}$.}
Given a real non-square matrix $\mat A$, the \hl{pseudoinverse} is $\mat A^+ \defn (\mat A\transp \mat A)^{-1}\mat A\transp$. 
Given some domain $\set X$,
we denote by $\set C^r(\set X)$ the set of $r$-times continuously differentiable functions over $\set X$, $r\geq1$,
and call a function $f$ smooth if $f\in\set C^\infty(\set D)$.
We denote the identity matrix by $\mat I$ {and the identity map by $\mathbf{id}$}. 
Given a single independent coordinate $t\in \mathbb R$ (here time), we call the map $\vec \gamma: \mathbb R \rightarrow \mathbb R^n$ a \hl{curve}. 
We denote the velocity of a curve $\vec \gamma$ at time $t$ by
$\dot{\vec \gamma}(t) \defn \frac{\mathrm d \gamma}{\mathrm d t} |_t$.
Given some $\set C^r$-map 
$\vec f: \mathbb R^n \rightarrow \mathbb R^m$
between Euclidean spaces $\mathbb R^n$ and $\mathbb R^m$, where $n,m\geq 1$, 
we denote the linear \hl{differential map} of $\vec f$ at $\vec x\in \R^n$ by $\mathrm{D} \vec f{(\vec x)}: \mathbb R^n \rightarrow \mathbb R^m$.
Given some point $\vec x \in \mathbb R^n$, we associate $\mathrm D \vec f{(\vec x)}$ with its Jacobian matrix $\mathrm D \vec f(\vec x)\in \mathbb R^{m\times n}$.
For a function $\vec f(\vec x, \vec y)$ of two arguments, we consider $\mathrm D_{\vec x} \vec f(\vec x, \vec y)$ and $\mathrm D_{\vec y} \vec f(\vec x, \vec y)$ separately.
An \hl{immersion} is a smooth map, $\vec h: \set Z \subseteq \R^{n_z} \rightarrow \set X \subseteq \R^{n_x}$, satisfying $\mathrm{rank}(\mathrm{D}_z \vec h^\dagger(\,\cdot\,)) = n_z$ (injective) everywhere in $\set Z$.

We consider embedded submanifolds of ambient space $\R^n$ characterized in terms of level sets.
The respective definition is intuitive and sufficiently general for our purposes.
\begin{definition}[Level set, \cite{Lee.Smooth}]
\label{def:level-set}
Let $\vec c:\set X \rightarrow \mathbb R^m$, $\set X \subseteq \mathbb R^n$ open, $1 \leq m < n$, be a $\set C^r$-mapping, $r\geq1$.
Denote by $\set{M} \subseteq \R^n$ the zero level set:
\begin{equation}
\label{eqn:zero-level-set}
	\set M = \{ \bm x \in \R^n : \bm c(\bm x)=\bm 0\} \,.
\end{equation}
Assume that $\mathcal{M}$ is non-empty and the Jacobian $\mathrm{D} \bm c(\bm x)\in \R^{m\times n}$ has constant rank $m$ at all $\bm x \in \set{M}$.
Then, $\set{M}$ is a manifold of class $\set C^r$, {\ie}, a differentiable manifold.
Moreover, $\set M$ has dimension $k = n - m$ and is embedded in $\R^n$, and therefore called an \hl{embedded $k$-submanifold} of $\R^n$.
By the definition of continuity, the preimage $\set{M} = \vec c^{-1}(\{\vec 0\})$ is closed in $\R^n$ and therefore $\set M$ is said to be properly embedded.
\end{definition}

We remark that the embedding is not an essential part of the definition, {\ie}, there exist more generic definitions that are independent of an ambient space \cite{Lee.Smooth}.
Therefore, we sometimes drop the terms \hl{embedded} or ``sub'' (manifold).
Note that $\R^n$ is a manifold itself.
By limiting ourselves to embedded submanifolds of open subsets $\set X \subseteq \R^n$ of Euclidean space $\R^n$ defined through level sets,
we circumvent geometric concepts like charts and tangent bundles.
We refer the interested reader to \cite{Lee.Smooth,Munkres.2018,Boumal.2023,Buchfink.2024}.
{Below, we always consider the case $m<n$, where $n=n_x$ (FOM dimension) and $k=n_z\geq 1$ (ROM dimension).}
{For notational and conceptual simplicity, we will focus on $\set C^\infty$ maps and $\set C^\infty$ manifolds. 
However, we remark that most methods are applicable to a less restrictive setup, 
such as $\set C^2$, which is occasionally also referred to as ``smooth'' in the literature.}

\section{Model order reduction problem}
\label{sec:mor-problem}
In this section, we present a general formulation of the MOR problem.
We are concerned with two types of non-autonomous dynamical systems {with inputs}.
We focus on nonlinear dynamical systems with external inputs,
described by a system of ordinary differential equations \abk{ODE}:
\begin{equation}
	\label{eqn:ode-control}
	\dot{\vec x}(t) = \vec f(\vec x(t), \vec u(t))\,,
\end{equation}
where $t\in[t_0,t_{\infty}]$ denotes time, 
$\vec x(t)\in \set X$ denote the differential states, $\set X \subseteq \R^{n_x}$, 
$\vec u(t)\in \set U$ {are the system inputs}, $\set U \subseteq \R^{n_u}$, 
and $\vec f: \set X \times \set U \rightarrow \R^{n_x}$ {is} smooth. 
We call $\set X$ the state space, $\set U$ the input space, and \eqref{eqn:ode-control} a state space model. 
{Notably, the input trajectory is generally unknown, \ie, $\vec u(t)$ are unknown time-variant parameters.}
The dimension $n_x$ is the \hl{model order}.

We further {consider} semi-explicit differential-algebraic systems of equations \abk{DAE} with external inputs:
\begin{subequations}
	\label{eqn:dae-nonautonomous}
	\begin{align}
	\dot{\vec x} (t) &= {\vec f}(\vec x(t), \vec y(t), \vec u(t)) \,,  \\
	\vec 0 &= {\vec g}(\vec x(t), \vec y(t), \vec u(t)) \,, \label{eq:aes}
	\end{align}
\end{subequations}
where $\vec x(t) \in {\set X} \subseteq \R^{n_x}$,
$\vec y(t) \in {\set Y} \subseteq \R^{n_y}$ are the algebraic variables,
and $\vec u(t) \in {\set U} \subseteq \R^{n_u}$ are the external inputs.
Let ${\set X}$, ${\set Y}$ and ${\set U}$ be open,
and 
${\vec f}: {\set X} \times {\set Y} \times {\set U} \rightarrow \R^{n_x}$
and
${\vec g}: {\set X} \times {\set Y} \times {\set U} \rightarrow \R^{n_y}$ be smooth.
We say that the DAE has \hl{index one} if $\mathrm D_{\vec y} {\vec g}(\hat{\vec x},\hat{\vec y},\hat{\vec u})$ has rank $n_y$ for any $(\hat{\vec x},\hat{\vec u}) \in \hat{\set X} \times \hat{\set U}$ on the solution set of ${\vec g}$ \cite{Brenan.1995,Parker.2022}. 

The class of semi-explicit DAEs covers a wide range of {process} models, where ODEs are often not sufficiently versatile. 
However, MOR methods are typically {presented in the} ODE {framework}.
Fortunately, this is not a problem because an index-one DAE \eqref{eqn:dae-nonautonomous} may be regarded as an ODE that lives on a submanifold of $\hat{\set X} \times \hat{\set Y} \times \hat{\set U}$ \cite{Rheinboldt.1984,Nijmeijer.1990}.
Applying Definition \ref{def:level-set} to \eqref{eq:aes} verifies this {point of view}.
Moreover, higher-index DAEs can usually be reduced to index-one systems \cite{Brenan.1995}.
Consequently, we can adopt ODE methods to index-one DAEs{, \cf~also} \cite{Loffler.1991,Hedengren.2005}. 
{Further works that explicitly deal with DAEs are, \eg, in \cite{Lee.1991,Sun.2005,Sjoberg.2007,Li.2023c,Vizzaccaro.2024}.}
For conciseness, we will only consider FOMs of the form \eqref{eqn:ode-control}.
However, the {MOR} methods can also be applied to \eqref{eqn:dae-nonautonomous} under the assumption of index-one. 

The second reason for introducing both \eqref{eqn:ode-control} and \eqref{eqn:dae-nonautonomous} is that some MOR methods reduce a high-order ODE to a low-order DAE, which we generally assume to be index-one.
Hence, we will frequently encounter ROMs of the form \eqref{eqn:dae-nonautonomous}.
Next, we formalize the MOR problem for nonlinear systems with external inputs:
\begin{problem}[MOR]
\label{pr:mor}
Given a FOM \eqref{eqn:ode-control},
find a low-order index-one DAE (the prototype of a ROM): 
\begin{subequations}
\label{eq:rom}
\begin{align}
    \dot{\vec z}(t) &= \vec f\ind{r}(\vec z(t), \vec u(t)), \label{eq:rom-dynamics}\\
    \vec 0 &= \vec h(\vec z(t), \vec x\ind{r}(t), \vec u(t))\,, \label{eq:rom-manifold}
    \end{align}
\end{subequations}
where differential states $\vec z(t)\in \set Z$, $\set Z \subseteq \R^{n_z}$, $n_z\ll n_x$,
algebraic variables $\vec x\ind{r}(t) \in \set X$ that approximate the FOM states $\vec x(t)$, 
and {where}
$\vec f\ind{r}: \set Z \times \set U \rightarrow \R^{n_z}$
and
$\vec h: \set Z \times \set X \times \set U \rightarrow \R^{n_x}$ are smooth. 
Let \eqref{eq:rom} be such that for any $\vec x(t_0)\in \set X${, $\vec u(t_0) \in \set U$,} and $\vec u(t){=\hat{\vec u}}\in \set U$, {$t > t_0$}, the prediction error{, $\varepsilon(t) \defn \norm{\vec x\ind{r}(t) - \vec x(t)}$, satisfies $\varepsilon(t)<\varepsilon^*$ for all $t > t_1$, where $t_1\geq t_0$ and $\varepsilon^*\geq 0$ are} sufficiently small.
\end{problem}
We call \eqref{eq:rom-dynamics} the \hl{latent dynamics} or \hl{internal dynamics} {of the ROM}.
For simplicity, we do not impose further requirements on the upper bound $\varepsilon^*$, {\eg}, {the} existence of a contraction mapping {\cite{Aulbach.1982,Khalil.2002}}.
{In general, a quickly decaying $\varepsilon(t)$ also depends on the ROM initialization strategy, where na\"ive minimization of $\lVert \vec x\ind{r}(t_0) - \vec x(t_0) \rVert_2$ may result in a suboptimal $\vec z(t_0)$ \cite{Roberts.1989,Otto.2022}.
}

{The dimension $n_z$ is the reduced model order.
A suitable value of $n_z$ is fundamentally system- and application-dependent and no universal rule can be given.
More specifically, the feasible degree of reduction depends on factors such as 
i) the type of dynamical system and its attractors (\eg, multiplicity, stability/hyperbolicity),
ii) the chosen reduction method,
iii) the required accuracy and computational costs of the ROM,
and
iv) the time-scales/frequencies and subset of state-space relevant to application (\eg, prediction horizon, time steps, bounded region on controllable manifold).
For example, some reduction methods provide specific requirements, \eg, on the slow/fast time-scale separation \cite{Haller.2017,Shen.2021}, thus limiting the feasible degree of reduction.
}

Next, we characterize the subset of $\set X$ on which the ROM \eqref{eq:rom} evolves.
To this end, we {first} examine the level set:
\begin{equation}
    \label{eq:level-set-dae}
    \set M \defn \{ (\hat{\vec x}, \hat{\vec z}, \hat{\vec u}) \in \set X \times \set Z \times \set U: \vec 0 = \vec h(\hat{\vec x}, \hat{\vec z}, \hat{\vec u}) \} .
\end{equation}
{Due to the index-one assumption in} Definition \ref{def:level-set}, $\set M$ constitutes a smooth embedded submanifold of $\set X \times \set Z \times \set U$, where $\dim \set M = n_z+n_u$.
In addition, when fixing $\hat{\vec u}\in \set U$, the level set:
\begin{equation}
\set M(\hat{\vec u}) \defn \{(\hat{\vec x}, \hat{\vec z}) \in \set X \times \set Z:  \vec 0 = \vec h(\hat{\vec x}, \hat{\vec z}, \hat{\vec u}) \}
\end{equation}
constitutes a smooth embedded $n_z$-submanifold of $\set X \times \set Z$, where $\set M(\hat{\vec u})\times \{\hat{\vec u}\}$ is a slice of $\set M$.
Finally, we denote the canonical projection of $\set M(\hat{\vec u})$ on $\set X$ by:
\begin{equation}
\label{eq:slow-manifold-u}
\set M_{\set X}(\hat{\vec u}) \defn \{\hat{\vec x} \in \set X :  \vec 0 = \vec h(\hat{\vec x}, \hat{\vec z}, \hat{\vec u}), \; \hat{\vec z} \in \set Z \}.
\end{equation}
To satisfy the requirements in Problem \ref{pr:mor}, we search for an invariant manifold $\set M$ having the property that $\vec f(\vec x(t), \vec u(t))$ is tangent to $\set M_{\set X}(\vec u(t))$ if $\vec x(t) \in \set M_{\set X}(\vec u(t))$.
Then, Eq.~\eqref{eq:rom} solves the invariance equations:
\begin{equation}
\label{eq:invariance}
\forall (\hat{\vec x}, \hat{\vec z}, \hat{\vec u}
) \in \set M :\;
\mathrm D_{\vec x} \vec h(\hat{\vec x},\hat{\vec z}, \hat{\vec u}
) \vec f(\hat{\vec x}, \hat{\vec u}
)
+
\mathrm D_{\vec z} \vec h(\hat{\vec x},\hat{\vec z}, \hat{\vec u}
) \vec f\ind{r}(\hat{\vec z}, \hat{\vec u}
) = \vec 0,
\end{equation}
which can be derived by differentiating \eqref{eq:rom-manifold} with respect to $t$ and inserting \eqref{eqn:ode-control} and \eqref{eq:rom-dynamics} for $\vec u(t)=\hat{\vec u}$ fixed.
Finally, notice that invariance is closely connected to conservation principles, \eg, the conservation of mass and energy \cite{Buchfink.2023,Zhang.2025}.

{
There exist various local and global approaches for determining invariant (sub)manifolds of nonlinear dynamical systems \cite{Krauskopf.2005,Touze.2021,Jain.2022}.
Within many of these approaches, the invariance equations \eqref{eq:invariance} play a key role.
For example, the parameterization method \cite{Cabre.2005,Haro.2006b,Haro.2016} and extensions \cite{Castelli.2015,vandenBerg.2016,Opreni.2023,Vizzaccaro.2024} have been proposed as a local intrusive method to construct invariant manifolds. 
If a rigorous solution of the invariance equations is impractical, the invariant manifold may be approximated based on system data (see Section \ref{sec:manifold-learning}).
Finally, projecting the FOM onto an invariant manifold enables a reduced-order representation valid on this subspace.
}

A fundamental difference to autonomous systems is that changes in $\vec u(t)$ can affect $\set M_{\set X}$.
For example, changing $\vec u(t)$ may rotate or shift the respective embedded $\set M_{\set X}$. 
If $\vec u(t)$ is piecewise constant, then $\set M_{\set X}$ may change its shape abruptly at the input grid points.
If $\vec u(t)$ is continuous, then $\set M_{\set X}$ may gradually deform over time.
When changing $\vec u(t)$ from $\hat{\vec u}^-$ to $\hat{\vec u}^+$ at time $t_1$, the prior $\vec x\ind{r}^- \defn \lim_{t\nearrow t_1}\vec x\ind{r}(t)$ may not be in $\set M_{\set X}(\hat{\vec u}^+)$.
{In the ROM,} this offset is immediately addressed by a projection of $\vec x\ind{r}^-$ onto $\set M_{\set X}(\hat{\vec u}^+)$, 
inherently defined by \eqref{eq:rom-manifold}.
{Accordingly, the ROM can exhibit a direct feedthrough from the inputs $\vec u(t)$ to the reconstructed states $\vec x\ind{r}(t)$, as indicated by \eqref{eq:rom-manifold}.}
This projection corresponds to the fast-time-scale response of the {FOM} to changes in $\vec u(t)$.
{Furthermore}, depending on $\vec h$, the associated projection {after a change in $\vec u(t)$} can introduce unwanted artifacts in the ROM response (see Section \ref{sec:spt}).

A common simplification of \eqref{eq:rom} is to use a one-to-one immersion
$\vec h^\dagger: \set Z \rightarrow \set X$,
so that $\set M_{\set X} = \{\vec h^\dagger(\hat{\vec z}): \hat{\vec z} \in \set Z\}$ 
and $\dot{\vec x}\ind{r}(t) \in \mathrm{col}(\mathrm D \vec h^\dagger(\vec z(t)))$ on $\set M$.
In this case, we have $\set M (\hat{\vec u}) \equiv \set M^*$, $\forall \hat{\vec u} \in \set U$, and $\set M = \set M^* \times \set U$ (trivial bundle), and {the invariance equations} \eqref{eq:invariance} become:
\begin{equation}
\label{eq:invariance-2}
\forall (\hat{\vec x}, \hat{\vec z}, \hat{\vec u}) \in \set M :\;
\vec f(\hat{\vec x}, \hat{\vec u}) 
-
\mathrm D_{\vec z} \vec h^\dagger(\hat{\vec z}) \vec f\ind{r}(\hat{\vec z}, \hat{\vec u}) = \vec 0.
\end{equation}

\vspace{-2ex}
\section{General-purpose reduction methods}
\label{sec:methods}
We review state-of-the-art reduction methods that are applicable to a wide range of systems, rather than developed for a specific type of process model.
Additionally, in Section \ref{sec:ml-control},
we develop an extension of the manifold-Galerkin method (reviewed in Section \ref{sec:manifold-learning}) to explicitly account for the dependence of the slow manifold on the inputs.
For the sake of brevity, we omit equations as far as possible and rather focus on an {informal} discussion.

\subsection{Intrusive general-purpose methods}
\label{sec:intrusive}
We begin with a review of intrusive MOR methods, which involve the FOM equations in the MOR procedure.
First, we discuss two classical intrusive approaches: 
proper orthogonal decomposition \abk{POD}
and singular perturbation theory \abk{SPT}. 
Afterwards, we present more recent extensions using machine learning, specifically manifold learning methods.

{The} POD belongs to the class of linear subspace methods,
{whereas} SPT constructs a nonlinear subspace.
Linear subspace methods are typically divided into methods based on singular values decomposition \abk{SVD} and Krylov subspace methods \cite{Antoulas.2001}.
Herein, we do not discuss Krylov subspace methods \cite{Bai.2002b}, 
which are major reduction techniques for linear systems,
but their extension to nonlinear MIMO systems is not straightforward.
We refer to \cite{Phillips.2003,Astolfi.2010} 
for some {recent} works towards this goal.
An extensive overview of MOR using Krylov subspace methods is given in \cite{Baur.2014}.

\subsubsection{POD-Galerkin method}
\label{sec:pod}
The POD, also known as Karhunen-Lo\'eve decomposition, 
is a method to extract a low-dimensional linear basis capturing the dominant behavior of a system \cite{Chatterjee.2000}.
Usually, the POD basis is determined through SVD of discrete state trajectory samples (\hl{snapshots}) from simulation studies.
Define the data matrix $\mat D\in \R^{n_x\times N}$, $N \geq n_x$, by stacking snapshots {$\vec x(t_k)$}, $k=1,2,...,N$.
Compute the mean $\bar{\vec x} \in \R^{n_x}$ and form the zero-centered matrix {$\mat X=[\vec x(t_1)-\bar{\vec x}, ..., \vec x(t_N)-\bar{\vec x}]$}.
Then, SVD provides: 
\begin{equation}
\label{eqn:svd}
\mat X = \mat U \mat \Sigma \mat V\transp{} \,,
\end{equation}
where {$\mat\Sigma\in\R^{n_x \times N}$} is a non-square diagonal matrix of ordered entries $\sigma_i$, which are the \hl{singular values}.
The orthogonal matrices $\mat U \in \R^{n_x \times n_x}$ and $\mat V \in \R^{N \times N}$ collect the corresponding left and right singular vectors, respectively.
The columns of $\mat U$ are also called \hl{principal components} of $\mat X$, since they are ordered by their relevance to represent the data $\mat X$.
As opposed to the eigendecomposition of $\mat X$, the SVD is well-conditioned and stable \cite{Antoulas.2005}.
However, SVD is sensitive to data scaling, so improper scaling can result in a poor ROM \cite{Chatterjee.2000}.

The \hl{POD-Galerkin method} consists of two main steps.
First, SVD provides the \hl{POD modes} $\mat U$.
Next, we specify the reduced order $n_z<n_x$ and subdivide $\mat U=[\mat U_1, \mat U_2]$, where $\mat U_1 \in \R^{n_x\times n_z}$.
Because $\mat U_1$ carries the dominant POD modes, we may truncate the mode $\mat U_2$ and use the approximation:
\begin{equation}
\label{eqn:pod-expansion}
\vec x(t_k) - \bar{\vec x} \approx \mat U_1 \vec z(t_k) ,\;
k=1,2,...,N ,
\end{equation}
where $\vec z(t_k) \in \R^{n_z}$ are lower-dimensional coordinates and $\bar{\vec x} \in \R^{n_x}$ is the mean used in the SVD above.
Eq.~\eqref{eqn:pod-expansion} corresponds to the optimal{, \ie,} \hl{proper}{,} approximation of the data $\mat X$ in the 2-norm \cite{Benner.2015}.
However, the optimality of $\mat U_1$ refers only to the {training} data and not to ROM prediction accuracy \cite{Rathinam.2003}.
The {reduced state space is defined as the} affine subspace, $\set M_{\set X} = \{\bar{\vec x} + \vec v: \vec v \in \mathrm{col}(\mat U_1)\}$.
Using \eqref{eqn:pod-expansion}, we perform a Galerkin projection of the FOM \eqref{eqn:ode-control} onto the basis $\mat U_1$ 
and obtain the ROM \cite{Rathinam.2003}: 
\begin{equation}
\label{eqn:rom-pod}
\begin{split}
\dot{\vec z}(t) &= \mat U_1\transp{} \vec f(\mat U_1 \vec z(t) + \bar{\vec x}, \vec u(t)),\\
\vec x\ind{r}(t) &= \mat U_1\vec z(t) + \bar{\vec x}.
\end{split}
\end{equation}
The POD is data-driven and therefore non-intrusive, but subsequent Galerkin projection is intrusive.
Combining data-driven reduced basis and physics-based FOM, the nonlinear ROM \eqref{eqn:rom-pod} is a hybrid model.
As the original map $\vec f$ is still contained in ROM \eqref{eqn:rom-pod}, an additional hyperreduction step may be required to enable real-time applications \cite{Bos.2004}.
In particular, the matrix $\mat U_1$ is usually dense, so a sparse model $\vec f$ may {even} lose its sparsity.

For {very high-dimensional} problems, sequential \hl{greedy} approaches are {computationally} cheaper {than SVD} \cite{Binev.2011,Hesthaven.2022}.
These methods minimize the $\infty$-norm of the error between reduced basis and snapshots rather than the least squares error or $2$-norm. 
The reduced basis is then generally non-orthogonal and a projection onto such a basis is a \hl{Petrov-Galerkin projection} \cite{Antoulas.2005}. 
Often, ROMs based on POD and greedy search {can} reach a similar accuracy \cite{Rozza.2008}.

\subsubsection{Singular perturbation theory}
\label{sec:spt}
Model reduction by SPT is a nonlinear subspace method based on time-scale separation of a FOM \cite{Kokotovic.1976,Fenichel.1979}.
Assume that the FOM \eqref{eqn:ode-control} is originally in \hl{singular perturbation canonical form} \cite{Khalil.2002}:
\begin{subequations}
\label{eqn:spt-original}
\begin{align}
\dot{\vec x}_1(t)&= \vec f_1(\vec x_1(t), \vec x_2(t), \vec u(t))\,,\label{eqn:spt-slow}
\\
\varepsilon \dot{\vec x}_2(t)&= \vec f_2(\vec x_1(t), \vec x_2(t), \vec u(t))\,, \label{eqn:spt-fast}
\end{align}
\end{subequations}
where $\varepsilon \ll 1$ and $\vec x_1(t) \in \R^{n_z}$ and $\vec x_2(t) \in \R^{n_x-n_z}$.
Eqs.~\eqref{eqn:spt-slow} and \eqref{eqn:spt-fast} collect the \hl{slow} and \hl{fast dynamics}, respectively.
We assume that exponential stability of the fast dynamics holds uniformly.

The key idea of perturbation theory is that the system response does not change significantly when $\varepsilon$ is perturbed.
When setting $\varepsilon = 0$ in \eqref{eqn:spt-original},
the fast dynamics degenerate into quasi-stationary equations, 
yielding a ROM defined by:
\begin{subequations}
	\label{eqn:spt-dae}\label{eqn:spt-rom}
	\begin{align}
	\dot{\vec x}_1(t)&= \vec f_1(\vec x_1(t), \vec x_2(t), \vec u(t)), \label{eqn:spt-de}
	\\
	\vec 0 &= \vec f_2(\vec x_1(t), \vec x_2(t), \vec u(t)).\label{eqn:spt-ae}
	\end{align}
\end{subequations}
Assuming that \eqref{eqn:spt-dae} is index-one, 
the corresponding slow manifold is given by:
\begin{equation}
\label{eqn:spt-manifold}
	\set M = \{(\hat{\vec x}_1, \hat{\vec x}_2, \hat{\vec u}) \in \set \R^{n_z} \times \R^{n_x-n_z} \times \R^{n_u} :
	\vec f_2(\hat{\vec x}_1, \hat{\vec x}_2, \hat{\vec u}) = \vec 0 \} \,.
\end{equation}
{Overall, the projection of \eqref{eqn:spt-original} onto $\set M$ is inherently defined via the degeneration.}

Typically, a process model is initially not in the form \eqref{eqn:spt-original}. Identifying the perturbation parameter $\varepsilon$ is {often} not obvious {and frequently based on human assumptions rather than system-theoretic arguments}.
{In many cases}, we seek a linear transformation,
$\vec x(t) = \mat T_1 \vec x_1(t) + \mat T_2 \vec x_2(t)$.
For chemical {process} systems, $\mat T_1$ and $\mat T_2$ are sometimes determined based on a time scale assumption on energy and mass transport \cite{Weischedel.1980,Roffel.2000}.
More rigorous approaches search for a subset of $\set X$ or $\set X \times \set U$ whose defect of invariance is sufficiently small \cite{Khalil.2002,Haro.2016,Jain.2022}.
{Further, note that implicit condensation \cite{Hollkamp.2008,Nicolaidou.2022} and static condensation \cite{Haller.2017} in mechanics are closely connected to SPT.
For example, the needed slow/fast time-scale separation for nonlinear vibrations was examined in \cite{Haller.2017,Shen.2021}.}

An important property of \eqref{eqn:spt-dae} is that the steady-state response remains exact. 
However, the degeneration of the fast dynamics occasionally introduces undesirable dynamic artifacts, especially a non-physical inverse response \cite{Benallou.1986,Ranzi.1988}.
{In other words, the degenerated finite-$\varepsilon$ dynamics have a substantial macroscopic effect when \eqref{eqn:spt-original} does not constitute a proper normal form \cite{Roberts.2008}.
Also,}
the appearance of an {inverse response} depends on the degree of reduction and the {determination} of $\mat T_1$ and $\mat T_2$ \cite{Horton.1991}.
In control, a sign-reversed model gain can degrade the closed-loop behavior of {a model-based controller} toward infeasibility or instability. 
However, a false {inverse response} is only problematic when appearing on the prediction time grid of the ROM application, 
where an extremely fast {inverse response} may be uncritical.

\subsubsection{{Residualization} method}
\label{sec:pod-res}
Closely related to the singular perturbation method is the \hl{POD residualization} \abk{POD-Res} or \hl{nonlinear-POD-Galerkin} method \cite{Foias.1988,Graham.1996}. 
Instead of Galerkin projection onto a linear subspace via POD truncation,
a nonlinear subspace ROM is obtained by considering the limit of infinitely fast (degenerate or \hl{slaved}) dynamics.
Let the POD modes $\mat U = [\mat U_1, \mat U_2, \mat U_3]$ so that $\mat U_1 \in \R^{n_x\times n_z}$ and $\mat U_2 \in \R^{n_x \times n_q}$,  
where $1 < n_q \leq n_x-n_z$.
Accordingly, we first extend \eqref{eqn:rom-pod} {and truncate $\mat U_3$ so that}:
\begin{subequations}
	\label{eqn:pod-res-conjugacy}
	\begin{align}
	\dot{\vec z}_1(t) &= \mat U_1\transp{} \vec f(\mat U_1 \vec z_1(t) + \mat U_2 \vec z_2(t) + \bar{\vec x}, \vec u(t)) \label{eqn:pod-res-leading-dynamics}
	\\
	\dot{\vec z}_2(t) &= \mat U_2\transp{} \vec f(\mat U_1 \vec z_1(t) + \mat U_2 \vec z_2(t) + \bar{\vec x}, \vec u(t)), \label{eqn:pod-res-residual-dynamics}\\
    \vec x\ind{r}(t) &= \mat U_1\vec z_1(t) + \mat U_2 \vec z_2(t) + \bar{\vec x} \,.
	\end{align}
\end{subequations}
Then, by residualization of \eqref{eqn:pod-res-residual-dynamics}, {\ie}, setting $\dot{\vec z}_2(t) = \vec 0$,
we obtain \cite{Shvartsman.1998}: 
\begin{subequations}
	\label{eqn:pod-residualization}
	\begin{align}
	\dot{\vec z}_1(t) &= \mat U_1\transp \vec f(\mat U_1 \vec z_1(t) + \mat U_2 \vec z_2(t) + \bar{\vec x}, \vec u(t)) \,, \label{eqn:pod-residual-leading}
	\\
	\vec 0 &= \mat U_2\transp{} \vec f(\mat U_1 \vec z_1(t) + \mat U_2 \vec z_2(t) + \bar{\vec x}, \vec u(t)) \,, \label{eqn:pod-residual-stationary}
	\\
	\vec x\ind{r}(t) &= \mat U_1\vec z_1(t) + \mat U_2 \vec z_2(t) + \bar{\vec x} \,.
	\end{align}
\end{subequations} 
Despite the POD-based partitioning of states in the first step, the residualization provides a \textit{nonlinear} subspace due to \eqref{eqn:pod-residual-stationary}.
In \cite{Schlegel.2002}, the authors compared POD-Galerkin and POD-Res, finding that low-order POD-Res ROMs are more accurate than POD-Galerkin ROMs but also more computationally expensive without hyperreduction.

When setting $n_q = n_x - n_z$, {\ie}, applying an exhaustive residualization with no truncation, POD-Res can be regarded as an SPT method.
This interpretation implies that the POD-based affine transformation divides the states into fast and slow dynamics.
In our experience, a certain degree of POD truncation is possible without a relevant loss of precision, yet enabling a significant improvement of numerical conditioning.
Therefore, we use $n_q < n_x - n_z$ below.
Finally, there exist alternatives to POD in the context of nonlinear Galerkin methods \cite{Marion.1989,Jauberteau.1990}, {\eg}, using wavelets \cite{Mahadevan.2000}.
In the next subsection, we discuss another nonlinear Galerkin variant based on machine learning, which circumvents the residualization step and is fully detached from a linear POD basis.

\subsubsection{Manifold-Galerkin method}
\label{sec:manifold-learning}
Applying machine learning to approximate a manifold from snapshot data is called \hl{manifold learning} \abk{MFL} or \hl{representation learning} \cite{Ma.2011,Szalai.2023}.
MFL encompasses a wide array of tools and techniques that can be grouped into (i) {learning methods}, including autoencoder networks \abk{AENs} 
\cite{Baldi.1989,Kramer.1992} 
and self-organizing maps \cite{Kohonen.1982}, 
(ii) spectral methods, such as classical PCA or kernel PCA \cite{Scholkopf.1997,Mika.1998},
locally linear embeddings \cite{Roweis.2000},
Laplacian eigenmaps \cite{Belkin.2003}, 
and diffusion maps \cite{Coifman.2006},
and (iii) probabilistic methods \cite{Ghojogh.2023}.
These methods can also be combined, {\eg}, 
{
classical PCA and quadratic decoding \cite{Barnett.2022,Geelen.2023} or
classical PCA and AENs \cite{Phillips.2021}}.

Not all MFL methods are suitable for MOR via Galerkin projection.
The subset of viable options includes AENs, self-organizing maps, and kernel PCA \cite{Lee.2020}.
Herein, we focus on MOR using AENs{, which are a prominent choice, \eg, \cite{Kashima.2016,Lee.2020,Gobat.2023}}.
We here consider \hl{undercomplete} AENs, where $n_z < n_x$.
First, the \hl{encoder} network $\vec \phi: \set X \rightarrow \R^{n_z}$, $\set X \subseteq \R^{n_x}$ open, projects the high-dimensional input vector $\vec x \in \set X$ to a lower-dimensional representation $\vec z\in \R^{n_z}$.
Often, $\vec z$ are called \hl{features} or \hl{latent variables}.
In the second step, the \hl{decoder} network $\vec \phi^\dagger: \set Z \rightarrow \R^{n_x}$, where $\set Z \subseteq \R^{n_z}$, reconstructs the high-dimensional vector from the features.
The images of the two mappings are $\vec \phi(\set X) \subseteq \set Z$ and $\vec \phi^\dagger(\set Z) \subseteq \set X$.
The standard unsupervised AEN training problem minimizes the mean squared reconstruction error, 
$\sum_{k=1}^N \lVert \vec x(t_k) - \vec \phi^\dagger \circ \vec \phi 
(\vec x(t_k)) \rVert_2^2$, given a data set $\set D = \{ \vec x(t_k) \}_{k=1}^N $ \cite{Vincent.2008}.
Assuming that $\vec \phi^\dagger$ is a one-to-one immersion, the subset $\set M_{\set X} \defn \{\vec \phi^\dagger(\hat{\vec z}): \hat{\vec z} \in \set Z\}$
is an $n_z$-dimensional embedded submanifold of $\set X$.
When using an {mean squared} training loss,
the composition $\vec \phi^\dagger \circ \vec \phi$ approximates an orthogonal projection from $\set X$ to $\set M_{\set X}$.

Intrusive MOR based on AENs is presented in \cite{Kashima.2016,Lee.2020} and refined{, \eg,} in \cite{Phillips.2021,Otto.2023}.
Here, we adopt the method of \cite{Lee.2020} and we comment on the refinements below.
Training an AEN on snapshots approximates the slow manifold by {$\set M_{\set X}\times \set U$}.
Differentiation of the approximation $\vec x(t) \approx \vec \phi^\dagger(\vec z(t))$ and insertion into \eqref{eqn:ode-control} yields:
\begin{equation}
\label{eqn:manifold-learning-ode-1}
\begin{split}
\mathrm{D}\vec \phi^\dagger(\vec z(t)) \dot{\vec z}(t) 
&= \vec f(\vec \phi^\dagger(\vec z(t)), \vec u(t)) \,,\\
\vec x\ind{r}(t) &= \vec \phi^\dagger(\vec z(t)) \,,
\end{split}
\end{equation}
where $\dot{\vec z}(t)\in \R^{n_z}$ and $\vec x\ind{r}(t)\in \R^{n_x}$.
Clearly, the map $\vec \phi^\dagger$ (and thus the activation function) need to be differentiable. 
To obtain a well-posed ROM, we multiply \eqref{eqn:manifold-learning-ode-1} by $\mathrm{D}\vec \phi^\dagger(\vec z(t))\transp$:
\begin{subequations}
\label{eq:lee-rom-0}
\begin{align}
\mathrm{D}\vec \phi^\dagger(\vec z(t))\transp \mathrm{D}\vec \phi^\dagger(\vec z(t))
\dot{\vec z}(t)
&= \mathrm{D}\vec \phi^\dagger(\vec z(t))\transp 
\vec f(\vec \phi^\dagger(\vec z(t)), \vec u(t)) \,,\\
\vec x\ind{r}(t) &= \vec \phi^\dagger(\vec z(t)) \,.
\end{align}
\end{subequations}
As a result, we obtain a \hl{manifold-Galerkin} projection of FOM \eqref{eqn:ode-control} onto $\set M_{\set X}$ \cite{Lee.2020}.
By applying the definition of the pseudoinverse, $\mathrm{D}\vec \phi^\dagger(\vec z(t))^+$, 
Eq.~\eqref{eq:lee-rom-0} can be finally brought into the form of \eqref{eq:rom}:
\begin{subequations}
\label{eqn:lee-rom}
\begin{align}
\dot{\vec z}(t)
&= \mathrm{D}\vec \phi^\dagger(\vec z(t))^+ 
\vec f(\vec \phi^\dagger(\vec z(t)), \vec u(t)) \,,\label{eqn:lee-rom-ode}\\
\vec 0 &= \vec x\ind{r}(t) - \vec \phi^\dagger(\vec z(t)) \,.
\end{align}
\end{subequations}
In practice, we employ \eqref{eq:lee-rom-0} rather than \eqref{eqn:lee-rom} to circumvent matrix inversion or SVD associated with pseudoinversion in the model.
Note that {both} \eqref{eqn:manifold-learning-ode-1} and \eqref{eqn:lee-rom} do not involve the encoding $\vec \phi$.
While $\vec \phi^\dagger (\vec z(t))$ is an embedding of $\set Z$ into $\set X$,
the map $\mathrm{D}\vec \phi^\dagger(\vec z(t))^+$ 
is a linear projection from $\R^{n_x}$ to $\R^{n_z}$ (more precisely from the tangent space to $\set X$ at $\vec \phi^\dagger(\vec z(t))$ onto the tangent space to $\set Z$ at $\vec z(t)$).
When using a linear AEN, then \eqref{eqn:lee-rom} reduces to \eqref{eqn:rom-pod}.
{Finally, we highlight the close connection between \eqref{eqn:lee-rom} and other approaches using a single parameterization/decoder map, \eg, \cite{Roberts.1989,Jain.2022}.}

{An alternative to pseudoinversion is to apply the left-inverse property,
$
\mathrm{D}\vec \phi(\vec \phi^\dagger (\,\cdot\,)) \circ \mathrm{D}\vec \phi^\dagger(\,\cdot\,)  = \mat I
$,
to \eqref{eqn:manifold-learning-ode-1}.
This approach is also valid for non-orthogonal (oblique) projections and yields \cite{Otto.2023}:
\begin{equation}
\label{eqn:otto-rom}
\dot{\vec z}(t)
= \mathrm{D}\vec \phi(\vec \phi^\dagger (\,\cdot\,)) 
\vec f(\vec \phi^\dagger(\vec z(t)), \vec u(t))
\end{equation}
instead of \eqref{eqn:lee-rom-ode}.
However, computing $\mathrm{D}\vec \phi(\vec \phi^\dagger (\,\cdot\,))$ can be expensive and the precise satisfaction of the left-inverse property is a crucial prerequisite of this method.
}

{Despite their intuitive structure and popularity in recent years, standard dense AENs suffer from limited applicability, poor scaling in the number of parameters and training costs, expensive model evaluation, moderate accuracy, and potential singularity issues.
We discuss these challenges next.}
{First}, the simultaneous strength and weakness of AENs is their feedforward structure, which renders AENs conceptually and numerically simple to deal with.
However, some manifolds cannot be globally represented by a single parameterization/decoder as provided by an AEN.
Consequently, AENs may not be applicable to some problems.
On the positive side, the slow manifold is represented by a feedforward structure rather than by a set of implicit equations{, see \eqref{eq:rom-manifold}}. 

{A second issue is that very high $n_x$ and $n_z$ cause an explosion in the parameters of dense encoder and decoder networks, resulting in intractable training and model evaluation.
Additionally, embedding}
{an AEN into a ROM} increases the number of {floating point operations of model evaluation} considerably.
For spatially distributed systems, computational improvements can be achieved by using convolutional instead of fully connected AENs \cite{Fresca.2022,Romor.2023}.
{Moreover, there exist more parameter-efficient approaches to realize the autoencoder structure, \eg, polynomials with 
compressed tensor coefficients have linear 
complexity scaling \cite{Grasedyck.2013,Szalai.2023}. 
On the other hand, we remark that the FOM equations, $\vec f$ in \eqref{eqn:lee-rom}, are often already computationally expensive, necessitating a hyperreduction step in any case.
Then, the AEN evaluation costs are secondary as the right-hand side of \eqref{eqn:lee-rom-ode} will be ultimately replaced by a cheaper function.
}

To improve ROM accuracy, the standard AEN training problem may be extended by 
{a robustness loss \cite{Zhou.2017b} as well as}
a velocity loss \cite{Otto.2023}{, which} we regard as a step towards \hl{physics-informed} model reduction.
{Furthermore, a standard AEN often faces injection errors, \ie, violating the identity $\vec \varphi \circ \vec \varphi^\dagger = \mathbf{id}$ or equivalently $\vec \varphi^\dagger \circ \vec \varphi \circ \vec \varphi^\dagger \circ \vec \varphi = \vec \varphi^\dagger \circ \vec \varphi$ (idempotence).
This identity can be included into the training problem \cite{Szalai.2023} or enforced structurally \cite{Otto.2023}.
Moreover, learning an invariant foliation instead of a single manifold may increase accuracy of the reduced space \cite{Szalai.2020,Szalai.2023}.}

{The AEN-based manifold-Galerkin method is sensitive to ill-conditioning of the Jacobian $\mathrm D \vec \phi^\dagger(\,\cdot\,)$.
Unfortunately, the Jacobian from a standard AEN training may be arbitrarily ill-conditioned. 
However, conditioning issues can be addressed by including a rank loss \cite{Nazari.2023} in the training or employing an invertible network structure possessing an exact (full-rank) Jacobian left-inverse \cite{Otto.2023}.}

Herein, we are particularly interested in systems with input.
Although \cite{Lee.2020,Otto.2023} consider input-affine or parametric systems, these works do not account for the effect of inputs or parameters on the slow manifold, {\cf}~\eqref{eq:slow-manifold-u}.
Instead, a trivial bundle $\set M = \set M_{\set X} \times \set U$ is constructed.
However, we expect this simplification to demand an unnecessarily high-dimensional slow manifold  {for some systems}.
Consequently, we employ the concepts from Section \ref{sec:mor-problem} to propose a respective extension below.

\subsubsection{Manifold-Galerkin method with inputs}
\label{sec:ml-control}
We extend the {machine learning} manifold-Galerkin method \cite{Lee.2020} to account for the {parametric} effect of system inputs on the slow manifold $\set M$ {and $\set M_{\set X}$}.
{This explicit parametric dependency has not been accounted for in the learning strategy proposed in \cite{Lee.2020} and related works (see Sections \ref{sec:manifold-learning} and \ref{sec:ml-nonintrusive}).}

Restricting ourselves to AENs, we are generally interested in maps of the form $\vec \varphi:\set X \times \set U \rightarrow \set Z \times \set U$ and $\vec \varphi^\dagger: \set Z \times \set U \rightarrow \set X \times \set U$.
The corresponding manifold is
{$\set M = \{(\hat{\vec x}, \hat{\vec z}, \hat{\vec u}) \in \set X \times \set Z \times \set U: \vec 0 = \hat{\vec x} - \vec \varphi^\dagger(\hat{\vec z}, \hat{\vec u})\}$},  where
{$\set M_{\set X} (\hat{\vec u}) = \vec \varphi^\dagger \circ \vec \varphi (\set X \times \{ \hat{\vec u}\})$}
and $\vec \varphi \circ \vec \varphi^\dagger = \mathbf{id}$. 
For convenience, we substructure $\vec \varphi^\dagger$ by stacking the maps $\vec \psi^\dagger: \set Z \times \set U \rightarrow \set X$ and $\mathbf{id}: \set U \rightarrow \set U$, so that
{$\set M_{\set X}(\hat{\vec u}) = \{ \hat{\vec x} \in \set X : 
\vec 0 = \hat{\vec x} - \vec \psi^\dagger(\hat{\vec z}, \hat{\vec u}),\, \hat{\vec z} \in \set Z \}$}.
Similarly, we substructure $\vec \varphi$ into the stacked maps $\vec \psi: \set X \times \set U \rightarrow \set Z$ and $\mathbf{id}: \set U \rightarrow \set U$.

{Given the above characterization of $\set M$,} Eq.~\eqref{eq:invariance} {becomes}:
\begin{equation}
\forall (\hat{\vec x}, \hat{\vec z}, \hat{\vec u}) \in \set M :\;
\vec f(\hat{\vec x}, \hat{\vec u})
-
\mathrm D_{\vec z} \vec \psi^\dagger(\hat{\vec z}, \hat{\vec u}) \vec f\ind{r}(\hat{\vec z}, \hat{\vec u}) = \vec 0 
{.}
\end{equation}
{To} guarantee that $\vec z(t)$ is not algebraically coupled to $\vec u(t)${, we  simplify $\vec \psi$ to $\vec \psi:\set X\rightarrow \set Z$.}
Consequently, {the encoding only depends on $\vec x(t)$.}
We apply an AEN training that minimizes the loss
{$\sum_{k=1}^N\lVert\vec x(t_k) - \vec \psi^\dagger(\vec \psi(\vec x(t_k)), \vec u(t_k)\rVert_2^2$}.
{Similarly to Section \ref{sec:manifold-learning}, this loss may be extended to incorporate system knowledge and reduce projection errors.}
We call our approach \hl{manifold learning with inputs} \abk{MFLu}.
Next, we project $\vec f$ onto {$\set M_{\set X}(\vec u(t))$}:
\begin{equation}
\mathrm{D}_{\vec z}\vec \psi^\dagger(\vec z(t), \vec u(t)) \, \dot{\vec z}(t) 
\vec f(\vec \psi^\dagger(\vec z(t), \vec u(t)), \vec u(t)).
\end{equation}
Finally, after left-multiplication by the pseudoinverse $\mathrm{D}_{\vec z}\vec \psi^\dagger(\vec z(t), \vec u(t))^+$,
we obtain the ROM:
\begin{equation}
\label{eq:bg-rom}
\begin{split}
\dot{\vec z}(t) 
&= \mathrm{D}_{\vec z}\vec \psi^\dagger(\vec z(t), \vec u(t))^+
\,
\vec f(\vec \psi^\dagger(\vec z(t), \vec u(t)), \vec u(t)) \,,\\
\vec x\ind{r}(t) &= \vec \psi^\dagger(\vec z(t), \vec u(t)) \,.
\end{split}
\end{equation}
This extension is related to studies on parametric model reduction; see \cite{Benner.2015,Benner.2017b} for an overview.
However, present works mostly deal with stationary parametric systems and use extensions of linear basis methods.
In particular, \cite{Ly.2001,BuiThanh.2003} combine a reduced POD basis and polynomial interpolation to build low-dimensional steady-state models of parametric problems.
Refinements of this approach are found in \cite{Mainini.2015,Hesthaven.2018}.
Reduction of parametric systems via manifold learning has also been investigated in the literature:
To reduce the dimensionality of steady parametric flow problems,
\cite{Franz.2014} present a strategy based on isomaps, 
and \cite{Diez.2021} apply kernel PCA.
In \cite{Holiday.2019}, diffusion maps are used to learn submanifolds of the input-output space for reducing the parameter dimension.
However, these works do not consider MOR of non-autonomous dynamical systems {with inputs}.
{Finally, we remark that MFLu is conceptually not restricted to AENs and may be adapted to other types of manifold learning.}

\subsection{Non-intrusive general-purpose methods} 
\label{sec:non-intrusive}
Besides the classical intrusive reduction methods, recent advances in {machine learning} and system identification have opened new avenues toward data-driven non-intrusive model reduction \cite{Klus.2018,Brunton.2020,Ghattas.2021}.
Importantly, non-intrusive reduction does not operate on the FOM equations and is therefore \hl{model-free}.
In addition, many non-intrusive methods combine MOR and hyperreduction in a single step. 
However, notice that non-intrusive reduction may still exploit some structural knowledge, {\eg}, knowing that the FOM is a polynomial model \cite{Kramer.2019}.
Finally, the generic and homogeneous structure of non-intrusive ROMs can be exploited in optimization \cite{Schweidtmann.2019,Schulze.2023}.
{Like Krylov subspace methods}, we disregard non-intrusive {rational approximation} \cite{Mayo.2007,Peherstorfer.2017,Scarciotti.2017,Nakatsukasa.2018}, as these methods are not (yet) applicable to nonlinear multi-input systems. 

\subsubsection{Dynamic mode decomposition}
\label{sec:dmd}
Among the conceptually most simple yet effective non-intrusive methods {is} dynamic mode decomposition \abk{DMD} \cite{Schmid.2010,Haller.2024}, {a} non-intrusive linear approach to construct linear ROMs, {\ie}, linear latent dynamics on a linear subspace.
{DMD} employ{s} SVD to identify and truncate the principal linear modes from trajectory snapshot data.
An extension of DMD to systems with inputs or controls \abk{DMDc} is presented in \cite{Proctor.2016}, providing a ROM:
\begin{equation}
\label{eq:dmdc-rom}
\begin{split}
    \vec z(k+1) &= \mat A\vec z(k) + \mat B \vec u(k),\\
    \vec x\ind{r}(k) &= \mat U_1 \vec z(k),
\end{split}
\end{equation}
where $\mat U_1 \in \R^{n_x \times n_z}$, $\mat A \in \R^{n_z\times n_z}$, and $\mat B \in \R^{n_z \times n_u}$.
We refer to \cite{Proctor.2016} for details on the computation of $\mat A$, $\mat B$, and $\mat U_1$.
For a stable linear FOM, \cite{Lu.2021b}
presents bounds on the prediction error, $\vec e(k) \defn \vec x\ind{r}(k) - \vec x(k)$, for $k>N$.
However, as discussed in \cite{Iacob.2024}, DMDc may {encounter inaccuracies when applied to} significantly nonlinear systems.

\subsubsection{Applied Koopman theory}
\label{sec:koopman}
An extension of the DMD algorithm was proposed in
\cite{Williams.2015}.
Their approach still provides a linear ROM, but the mathematical procedure to construct the dynamics involves a prior nonlinear lifting step based on Koopman theory.
Koopman theory \cite{Mezic.2005,Otto.2021}
postulates that a nonlinear dynamical system possesses a linear operator representation when lifted
to a generally infinite-dimensional space of observables $g: \set X \rightarrow \mathbb{C}$.
Constructing a finite-dimensional or even low-order Koopman model by means of (E)DMD corresponds to a Galerkin projection of the Koopman operator onto a set of basis functions $g_i$, collected in {a vector} $\vec g$.
Hence, despite being a linear method,  
DMD is applicable to nonlinear systems, when substituting state snapshots $\vec x^{(k)}$ by a sufficiently rich set of embedded snapshots $\vec g(\vec x^{(k)})$.
Common choices of real-valued observables include Hermite polynomials,
radial basis functions, sparse regression, and ANNs \cite{Williams.2015,Li.2017}.
More recently, kernel methods have gained attention due to their strong approximation capabilities \cite{Das.2020,Heas.2025,Kohne.2025}. 

Notably, a Koopman model does not necessarily predict states but only deals with observables.
However, we can adapt the Koopman framework to state prediction by demanding that $\vec x(t)$ can be inferred from $\vec g(\vec x(t))$.
To this end, a common assumption is that $\vec x(t)$ can be linearly reconstructed from the nonlinear observables.
For example, \cite{Korda.2018} and follow-ups \cite{Narasingam.2019,Hara.2023,Hao.2024} 
use {extended DMDc}:
\begin{equation}
\label{eq:edmdc}
\begin{split}
    \dot{\vec z}(t) &= \mat A \vec z(t) + \mat B \vec u(t),\\
    \vec z(t_0) &= \vec g(\vec x(t_0)),\\
    \vec x(t) &= \mat C \vec z(t),
    \end{split}
\end{equation}
where $\vec z(t) \in \R^{n_z}$ are the Koopman states, 
$\vec g: \set X \rightarrow \R^{n_z}$ is the nonlinear observable map,
and $\mat A \in \R^{n_z \times n_z}$, $\mat B \in \R^{n_z \times n_u}$, $\mat C \in \R^{n_x \times n_z}$.
A ROM is obtained when setting $n_z < n_x$.

\begin{figure}[t!!]
	\centering
	\includegraphics[width=0.65\linewidth]{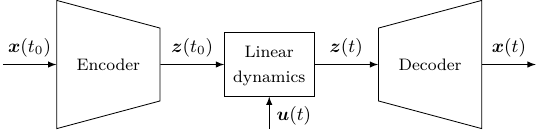}
	\caption{Model structure of the Koopman-Wiener ROM.}
	\label{fig:koopman_wiener_linear}
\end{figure}

In the past years, Koopman theory has become a popular approach for both model bilinearization \cite{Williams.2016,Surana.2016,Peitz.2020} 
and non-intrusive MOR \cite{Otto.2019,Peitz.2019}.
For autonomous systems, \cite{Lusch.2018} relaxes the assumption of a global linear projection and proposed to combine linear dynamics and nonlinear reconstruction by means of a decoder network $\vec g^\dagger$.
In \cite{Schulze.2022a}, we have shown that this strategy is also valid for input-affine systems.
Under certain assumptions on the observables $\vec g${,} see  \cite{Schulze.2022a}, the respective ROM takes the form:
\begin{equation}
\label{eq:KW}
\begin{split}
    \dot{\vec z}(t) &= \mat A \vec z(t) + \mat B \vec u(t),\\
    \vec z(t_0) &= \vec g(\vec x(t_0)),\\
    \vec x\ind{r}(t) &= \vec g^\dagger(\vec z(t)),
    \end{split}
\end{equation}
which is visualized by Fig.~\ref{fig:koopman_wiener_linear}.
Notice that the linear dynamics in \eqref{eq:KW} is a special case of the bilinear dynamics derived in \cite{Schulze.2022a}.
For non-oscillating systems, we may choose a diagonal $\mat A$ and thereby reduce the trainable parameters.
The combination of linear dynamics and nonlinear static map has \hl{Wiener structure} and we term \eqref{eq:KW} a \hl{Koopman-Wiener} \abk{KW} model.
In contrast to \eqref{eqn:lee-rom}, we do not only project the nonlinear dynamics onto some subspace of $\set X$, but we additionally demand the projection $\vec g$ to be linearizing.
Consequently, the learning problem is more complex than standard MFL.
Realizing the KW framework is most straightforward in a non-intrusive fashion, where a training {problem} combines reconstruction, prediction, and regularization loss terms \cite{Schulze.2022a}.
Finally, the flexibility of {machine learning} enables modifications of the KW structure.
For example, the encoder may serve as a state estimator \cite{Schulze.2022b}.

\subsubsection{Lift-and-project methods}
\label{sec:lift-and-learn}
Lift-and-project methods expand the FOM into an even higher-dimensional canonical form before projecting this representation to a low-dimensional subspace.
The lifting step usually targets a linear, bilinear, or quadratic representation of the FOM.
{Notably,} this extra step enables (i) the use of linear or {quadratic/polynomial} MOR methods, and (ii) circumvents hyperreduction due to mathematically simpler ROMs.
Methods such as Koopman lifting \cite{Korda.2020b}, 
Carleman bilinearization \cite{Steeb.1980}, 
piecewise linearization \cite{Rewienski.2003},
{variational analysis via} Taylor expansion \cite{Phillips.2003,Feng.2004}, 
{and quadratization \cite{Gu.2011,Hemery.2020,Bychkov.2024}}
lay the theoretical foundation of lift-and-project methods.
{A detailed overview is given in \cite{Baur.2014} and software to automate the lifting was recently presented in \cite{Bychkov.2024,Hemery.2021,Fages.2025}.}

{In cases where} intrusive variants of lift-and-project methods are mathematically involved {or too} computationally expensive, 
non-intrusive implementations are convenient to circumvent lifting the FOM. 
For example, the studies \cite{Kramer.2019,Qian.2020}
presented a non-intrusive lifting method for polynomial FOMs, where snapshot data is lifted and POD-projected to learn a quadratic ROM.
Therein, the model regression of the quadratic ROM is called \hl{operator inference} \cite{Peherstorfer.2016,Benner.2020}.
{Extensions of operator inference to polynomial manifolds and ROMs are found, \eg, in \cite{Geelen.2023,Geelen.2024}.}
{Also, the non-intrusive (extended) DMDc and KW methods can be regarded as a variant of operator inference, where the latent dynamics are restricted to a linear or bilinear form.}

\subsubsection{Reduced basis and manifold learning methods}
\label{sec:ml-nonintrusive}
An alternative non-intrusive reduction approach is to combine reduced basis methods or manifold learning with system identification methods.
For example, given POD-reduced snapshots, the latent ROM dynamics may be learned using recurrent neural networks \cite{Wang.2018b}, 
temporal convolutional ANNs \cite{Wu.2020},
or feedforward ANN predictors \cite{Gao.2020}.
Analogously, given AEN-compressed snapshots,
an ROM can be obtained, {\eg}, using {ANNs} \cite{Wiewel.2019} or Hammerstein-Wiener models \cite{Tsay.2020}.
These approaches are non-intrusive counterparts to \eqref{eqn:lee-rom}.
{Non-intrusive operator inference for quadratic manifolds was presented in \cite{Geelen.2023}, where the authors infer a quadratic ROM from POD-projected snapshot data.
This approach was extended to polynomial manifolds and ROMs in \cite{Geelen.2024}.}

The above methods are two-stage procedures because the slow manifold is constructed independently of the latent dynamics.
However, this separation may introduce a bottleneck, {\ie}, the latent variables may not be suitable latent states, depending on the type of predictor.
Interestingly, a similar bottleneck is also present in intrusive model reduction, as highlighted in \cite{Otto.2022,Otto.2023}.
Non-intrusive single-stage approaches address this issue by  learning slow manifold and latent dynamics simultaneously.
Conveniently, the KW method reviewed above inherently follows this strategy \cite{Schulze.2022a}.
In a similar but empirical setup, the works \cite{Watter.2015,Goroshin.2015} use AENs and linear time-invariant dynamics to learn stochastic ROMs.
{A combination of AEN and quadratic dynamics is used in \cite{Goyal.2024}.}
Alternative approaches combine MFL and {recurrent ANNs}
\cite{Gonzalez.2018} {or} feedforward ANNs \cite{Assael.2015,Masti.2021,Fresca.2021,Gobat.2023} \abk{termed MFL-ANN below},
{as well as} MFL and sparse identification of latent dynamics \cite{Conti.2023}.

\subsection{Discussion and comparison}
\label{sec:comparison}
We present a comparison of the reviewed methods with respect to their nonlinear reduction properties in 
Table \ref{tab:comparison}.
First, we distinguish methods by their intrusive character (\textbf{Intr}).
Intrusive MOR methods do not inherently perform hyperreduction (\textbf{Hr}) and thus require additional model simplification efforts for real-time applications.
In contrast, non-intrusive methods use ROM structures that are typically (by design) much cheaper to evaluate so that a separate hyperreduction step is not needed.

Next, we distinguish whether the methods are snapshot-based (\textbf{Snap}).
This categorization reveals another practical limitation{, namely the ability to deal with a large number of inputs (\textbf{Nu})}.
Snapshot-based methods usually face the curse of dimensionality in $n_u$,
since an independent excitation of inputs is commonly used to adequately cover the relevant state-space regions.
As a consequence, snapshot-based methods are less straightforward to apply to systems with many inputs, {\ie}, $n_u \gg 1$.
Hence, for systems with many independent inputs $\vec u(t)$, snapshot-free intrusive methods are the most suitable.

Despite the curse of dimensionality {in data sampling}, 
input-response snapshots contain valuable information about controllability.
As discussed in \cite{Willcox.2002}, when the snapshots are created via broad excitation of the system through the inputs, these data sample the controllable subspace.
As a consequence, building a ROM based on input-response data inherently approximates the controllable subspace and eliminates the uncontrollable states (\textbf{Ctr}).

\begin{table}[t!]
    \centering
    \caption{
    Comparison of nonlinear MOR methods that preserve state information.
    \textbf{Intr}: intrusive method,
    \textbf{Snap}: snapshot-based method,
    \textbf{Manif}: type of submanifold,
    \textbf{Lat}: type of latent dynamics,
    \textbf{Hr}: hyperreduction features,
    \textbf{Nu}: suitable for many inputs ($n_u \gg 1$),
    \textbf{Ctr}: approximates controllable subspace, 
    \textbf{Stab}: preserves stability.
    {\yes: yes,}
    (\yes): practical limitations,
    {\no: no.}
    }
    \label{tab:comparison}
    \resizebox{\textwidth}{!}{
    \begin{tabular}{lll|cccc|cccc}
    \toprule
    & \textbf{Method} & \textbf{Section} & \textbf{Intr}  & \textbf{Snap} & 
    \textbf{Manif} & \textbf{Lat} & 
    \textbf{Hr} & \textbf{Nu} & \textbf{Ctr} & 
    \textbf{Stab}\\ 
    \midrule
    a) & POD-Galerkin &  \ref{sec:pod}  
    & \yes & \yes & Lin.& Nonlin. & 
    \no & (\yes) & \yes & {(\yes)}  \\
    b) & SPT & \ref{sec:spt} 
    & \yes & \no &  Nonlin. & Nonlin. & 
    \no & \yes & \no & \yes \\
    c) & POD-Res. & \ref{sec:pod-res} 
    & \yes & \yes & Nonlin. & Nonlin. & 
    \no & (\yes) & \yes & \yes \\
    d) & MFL-Galerkin & \ref{sec:manifold-learning} 
    & \yes & \yes & Nonlin. & Nonlin. & \no & (\yes) & \yes & \no \\
    e) & MFLu-Galerkin & \ref{sec:ml-control} 
    & \yes & \yes & Nonlin. & Nonlin. & \no & (\yes) & \yes & \no \\
    f) & {(Extended)} DMDc & \ref{sec:dmd}/\hyperref[sec:koopman]{2} 
    & \no & \yes & Lin. & Lin./Bilin. &
    \yes & (\yes) & \yes & {\yes}   \\
    g) & KW & \ref{sec:koopman} 
    & \no & \yes & Nonlin. & Lin./Bilin. & 
    \yes & (\yes) &\yes & \yes   \\
    h) & {Operator inference} & \ref{sec:lift-and-learn} 
    & \no & \yes & {Polyn.} & {Polyn.} & 
    \yes & (\yes) &\yes & \yes  \\
    i) & POD-ANN & \ref{sec:ml-nonintrusive} 
    & \no & \yes & Lin. & Nonlin. & 
    \yes & (\yes) & \yes&\no   \\
    j) & MFL-ANN & \ref{sec:ml-nonintrusive} 
    & \no & \yes & Nonlin. & Nonlin. & 
    \yes & (\yes) & \yes & \no  \\
    \bottomrule
\end{tabular}
}
\end{table}

Next, we distinguish between methods that construct a linear (or affine) subspace versus a nonlinear manifold (\textbf{Manif}).
As discussed in Section \ref{sec:intro}, using a linear subspace can limit the possible degree of reduction.
On the other hand, computing a reduced basis by means of POD is numerically robust and computationally cheaper than MFL.
The effort of a system-theoretic reduction, such as SPT, depends on the system knowledge and availability of heuristics. 
Overall, there is a trade-off between reduced order and reduction effort.

Lastly, we distinguish the reduction methods by the type of latent dynamics (\textbf{Lat}), see \eqref{eq:rom-dynamics}.
In general, the ROM of a nonlinear FOM is also nonlinear.
However, DMDc and KW reduction methods construct linear latent dynamics, whereby the nonlinear ROM is simplified and \hl{structurally regularized}.
In other words, a ROM with linear latent dynamics cannot falsely exhibit strong nonlinear phenomena such as chaos or state multiplicity \cite{Pearson.2003}.
At the same time, these ROMs are limited to mildly nonlinear systems or short prediction horizons.
However, a ROM with linear or bilinear latent dynamics is oftentimes sufficient, where more complex structures, {\eg}, MFL-ANN, bear a higher risk of overfitting and undesired dynamical behavior.

{Only a subgroup of} nonlinear model reduction methods provide a-priori guarantees on stability preservation (\textbf{Stab}).
These guarantees are established by system-theoretic arguments (SPT, AGG, POD-Res.)~or through constrained learning problems ({DMD}, KW, {and operator inference}).
{In reduced basis methods, stability preservation may be enabled through variable transformation or modification of the inner product \cite{Rowley.2004,Barone.2009}.}
Other methods require a careful a-posteriori investigation of {ROM} stability \cite{Benner.2015,Hesthaven.2022}. 
{We also mention methods developed to incorporate physics priors,
particularly for
Hamiltonian \cite{Afkham.2017,Sharma.2022,Gruber.2023,Buchfink.2023},
port-Hamiltonian \cite{Gugercin.2012,Mehrmann.2023,Geng.2025},
and Lagrangian systems \cite{Carlberg.2015,Friedl.2024}.
These formalisms support the construction of ROMs that respect energy and mass conservation and preserve stability \cite{Buchfink.2023,Lepri.2023}.
Here, the port-Hamiltonian framework is particularly suitable for dynamical systems with inputs or interconnected process subsystems.}

Regarding the ROM error, $\vec e(t) \defn \vec x\ind{r}(t) - \vec x(t)$, {many publications on} MOR techniques do not {derive} universal error bounds but {perform} a-posteriori ROM validation.
{Rigorous error bounds are presented,} \eg, in {\cite{Aulbach.1982,Khalil.2002,Besselink.2013b,Shakib.2023,Kawano.2026}}.
{The accuracy of a ROM is directly coupled to modal closure and memory effects \cite{Ahmed.2021}.
Hence, methods for ROM closure modeling, \eg, the Mori-Zwanzig formalism \cite{Gouasmi.2017}, support tighter error bounds. }

\section{Reduced modeling of chemical processes}
\label{sec:pse}
We now review specialized MOR methods developed in the {process systems engineering} field as well as applications of general-purpose methods to {such} problems.
In Section \ref{sec:pse-methods}, we review three specializations of MOR methods from the previous section, namely
(i) compartment modeling, 
(ii) model aggregation,
and (iii) collocation method.
Afterwards, in Section \ref{sec:pse-applications}, we discuss the application of MOR to three very common unit operations: (i) distillation columns, (ii) heat exchangers \abk{HX}, and (iii) reactors.
Again, we omit equations as far as possible and rather focus on an {informal} discussion.
Besides deterministic reduction, 
reduced modeling is often achieved based on knowledge or intuition through simplifying assumptions in the modeling process \cite{Esche.2014}.
We also comment on these aspects.

\subsection{Problem-specific approaches}
\label{sec:pse-methods}
\subsubsection{Compartment modeling}
\label{sec:compartment}
The key concept in compartment modeling \abk{COMP} is the definition of interconnected functional subsystems, termed \hl{compartments}, between the local microscale and the global system scale \cite{Jourdan.2019}.
When applied for MOR, the introduction of compartments is often motivated by a time scale assumption, where local phenomena inside a compartment are assumed to be much faster than the overall response of the compartment \cite{Benallou.1986}.
Given a discretized model of a conservative distributed system,
a compartment is a cluster of neighboring cells, and the compartment equations are obtained by superposition of the extensive balance equations, {\eg}, absolute enthalpy or total molar holdup, of these cells.
In terms of intensive quantities, {\eg}, molar enthalpy and molar fraction, this superposition corresponds to a mass-weighted averaging.

By applying the above time scale assumption, the local dynamics of cells within a compartment are degenerated to quasi-stationary equations, and we obtain a ROM.
In addition, replacing quasi-stationary equations by shortcut methods, averaging rules, or regression models realizes a hyperreduction.
COMP is a heuristic approach to transform an FOM into Eqs.~\eqref{eqn:spt-original} and \eqref{eqn:spt-rom}.
Hence, we regard COMP as a heuristic SPT method,
inheriting properties such as the potential appearance of a non-physical {inverse response}.

\subsubsection{Model aggregation}
\label{sec:aggregation}
In order to eliminate the potential {inverse response} in COMP ROMs and improve model sparsity,  \cite{Levine.1991} presents the \hl{aggregation method} \abk{AGG}.
AGG is a heuristic modification of SPT for 1D distributed systems. 
Instead of transforming the FOM into the canonical form \eqref{eqn:spt-original},
the authors multiply (only) the left-hand side of the FOM by a matrix $\tilde{\mat H} =\mathrm{diag}(\tilde{h}_1, \tilde{h}_2, ..., \tilde{h}_{n_x})$, $\tilde{h}_{i} \geq 0$.
Thereby, the time constants of individual stage dynamics are artificially increased or reduced.
$\tilde{\mat H}$ is a hyperparameter and commonly {selected such that} $\mathrm{trace}(\tilde{\mat H}) \approx n_x$.
Let $\mat \Omega$ be the $n_x \times n_x$ permutation matrix such that 
$\mat \Omega \tilde{\mat H} = \big[ \begin{smallmatrix}
    \mat H & \mat 0 \\ \mat 0 & \mat 0 
\end{smallmatrix} \big]$,
where $\mat H= \mathrm{diag}(h_1,h_2,...,h_{n_z})$, $h_i > 1$.
The resulting ROM reads:
\begin{subequations}
    \label{eqn:aggr-rom}
	\begin{align}
	\mat H \dot{\vec x}_1(t)&= \vec f_1(\vec x_1(t), \vec x_2(t), \vec u(t))\,,
	\\
	\vec 0 &= \vec f_2(\vec x_1(t), \vec x_2(t), \vec u(t))\,.
    \label{eq:agg-ae}
	\end{align}
\end{subequations}
A systematic comparison of AGG to COMP is given in \cite{Linhart.2010}.
By contrasting \eqref{eqn:spt-rom} and \eqref{eqn:aggr-rom}, we notice that the manifolds on which the ROMs evolve are identical if $\vec f_2$ coincide.
Hence, both methods may realize a different ROM living on the same manifold $\set M$.

\subsubsection{Collocation method}
Motivated by the collocation method for solving {ordinary and partial differential equations} \cite{Hairer.1993,Franke.1998},
the studies \cite{Wong.1980,Cho.1983b} adapt the collocation method to MOR of spatially discrete problems.
The underlying assumption is a polynomial coherence of the differential variables of the FOM.
Then, a ROM is obtained by combining polynomial interpolation and a number of discrete cells (collocation points), {\eg}, finite volume elements or column stages. 
However, the collocation points do not necessarily correspond to cells in the FOM.
Intuitively, the number and position of collocation points depend on the system, {\ie}, whether a low-order polynomial is able to capture the underlying pattern{s characterizing the} slow manifold.
The positions of the collocation points are typically chosen a-priori as the roots of Legendre, Hahn, or Jacobi polynomials and not further adapted online \cite{Stewart.1985,Swartz.1986}.
In other words, the spatial collocation points are determined once and for all, {\ie}, do not change over time.

\subsection{Applications}
\label{sec:pse-applications}
\subsubsection{Distillation columns}
\label{sec:distillation-roms}
Detailed distillation column models typically include \hl{stage-by-stage} differential energy, mass, and possibly momentum balances \cite{Howard.1970}. 
Common simplifying assumptions are ideally mixed stages with no diffusion effects, thermodynamic vapor-liquid equilibrium, and quasi-stationary pressure and enthalpy dynamics of both phases \cite{Ranzi.1988,Roffel.2000}.
These assumptions rely on time-scale assumptions and may be regarded as a heuristic application of SPT, rendering such mechanistic models already reduced.
However, these models are often too expensive for online applications and therefore require MOR \cite{Abdullah.2007,Schulze.2021}.

Early applications of MOR to columns are found in \cite{Wong.1980,Cho.1983b} for COL and  \cite{Gilles.1980} for COMP.
As mentioned above, when applying COL, the collocation points do not necessarily correspond to physical separation stages, {\ie}, can be pseudo-stages.
Moreover, constructing COL models based on a log-transformed FOM can improve model accuracy \cite{Cao.2016}.

Using SPT, \cite{Benallou.1986} derives a widely applicable COMP model form and shows a higher accuracy over COL.
The COMP ROMs therein combine dynamic compartment balances and quasi-stationary stage models.
The separation stage whose differential balances are replaced by compartment balances is termed \hl{sensitivity stage}.
The appearance of a non-physical {inverse response} depends on the number of compartments and the position of the sensitivity stages \cite{Horton.1991,Huang.1993}.
However, avoiding an {inverse response} may limit the degree of order reduction.
On the other hand, the particular structure of COMP models,
featuring large sets of neighboring quasi-stationary stage models, 
enables hyperreduction via steady-state shortcut methods, {\eg}, FUG shortcut \cite{Diwekar.1994} or group methods \cite{Kamath.2010,Ecker.2019}. 
Typically, these shortcuts have fewer parameters than universal regressors ({\eg}, ANNs) and are less prone to overfitting.
On the other hand, ANNs can reach almost arbitrary precision \cite{Schafer.2019}.

In steady operation, a distillation column exhibits a set of coherent composition patterns, which can be analytically derived under simplifying assumptions \cite{Marquardt.1986,Hwang.1989}.
These analytical solutions are wave-shaped profiles, and therefore called \hl{nonlinear wave propagation model}.
{Ref.~}\cite{Kienle.2000} proposed a semi-empirical ROM for multi-species distillation columns, by combining {wave propagation} equations, global column balances, and a boundary equilibrium assumption.
The resulting ROM is a compartment model (a non-physical {inverse response} is also observed here).
In \cite{CaspariWave}, this ROM was extended to account for effects due to non-constant holdup.

The AGG method was originally developed for distillation columns \cite{Levine.1991}.
Like COMP models, AGG ROMs involve clusters of quasi-stationary trays that can be simplified using shortcut models.
Different from COMP, the AGG ROM states may be associated with actual stages in a distillation column.
While avoiding a non-physical {inverse response}, the AGG {predictions} usually converge more slowly to the FOM states after a change of system inputs \cite{Linhart.2009}.

Besides the specialized MOR methods, general-purpose MOR methods have been applied to distillation column models, {\eg}, using
POD-Galerkin \cite{Hahn.2002c,Hedengren.2005b},
standard SPT \cite{Kumar.2003},
KW \cite{Schulze.2022b,Schulze.2023}, 
and DMDc \cite{Qian.2023}.
However, as opposed to these general-purpose methods,
the specialized methods, COMP{,} {nonlinear wave propagation}, AGG, and COL, build on the intrinsic system structure, providing generic low-order models that can be written (almost) independently of a FOM.
In particular, these specialized {ROMs} can be pre-configured for generic columns with similar stage models and in many cases steady-state data is sufficient to {customize} these ROMs for a specific column.
Moreover, COMP and AGG may access an extra set of hyperreduction methods in the form of steady-state shortcuts as described above.

\subsubsection{Heat exchangers}
HX are another type of distributed systems that are commonly associated with high-order FOMs.
For industrial processes, the 3D discretized model of a multi-stream HX can reach an order of over a million states \cite{Haider.2020}.
Clearly, such fine-grained models are prohibitive for real-time applications.

As an alternative to MOR, simplifying assumptions enable a heuristic degeneration or lumping of differential equations.
Common assumptions are (i) a time scale separation between energy dynamics of fluids and walls and
(ii) for parallel flow HX, a uniform \hl{common wall} temperature in each cross section perpendicular to the principal flow axis \cite{PiconNunez.2002}.
Common examples of lumped parameter models are $\varepsilon$-NTU, P-NTU and LMTD \cite{Shah.2003}.
However, these are typically only valid at stationary points and under simplified assumptions on geometry, heat transfer, and thermodynamics.

Under reasonable simplifying assumptions, such as constant heat capacities and linear heat transfer correlations, the HX FOMs are (quasi-)linear or bilinear \cite{Luo.2002}.
In that case, linear or bilinear MOR may be applied \cite{Xia.1991}, which is less involved than general nonlinear MOR.
Moreover, when including dynamical HX submodels in a process model,
the heat transfer phenomena often have significantly smaller time constants than mass conversion and mass transfer problems \cite{Baldea.2006}.
Under these assumptions, a dynamical HX model may fully degenerate into a quasi-stationary model.

Applications of general-purpose MOR methods to HX include POD-Galerkin \cite{Xu.2018,Christ.2018,Ma.2021},
DMD \cite{Xu.2024},
and POD combined with ANN {predictors} \cite{Li.2024}.
Furthermore, both classical collocation and Galerkin projection are standard methods to approximate low-order solutions of {partial differential equations} \cite{Hansen.1974,Shah.2003}. 
In addition, all three specialized approaches (COMP, AGG, COL) have been applied to spatially discretized dynamical models of 1D distributed HX.
Specifically, \cite{Cao.2016,Quarshie.2023} use the collocation method to reduce finite volume models of HXs.
Lumped COMP models of HX are presented in
\cite{Mathisen.1994,ZavalaRio.2007,Michel.2013}.
Also, the reduced modeling of HX with phase change (evaporators or condensers) is often accomplished using the \hl{moving boundary model} \cite{Jensen.2002}, which is a COMP model.
The {moving boundary model} uses compartments with flexible boundaries to account for spatially moving phase regions and can exhibit a non-physical {inverse response} \cite{Vaupel.2019}.
Finally, AGG is applied to HX in \cite{Linhart.2011}.

Similar to separation columns, there exist steady-state shortcut models and analytic solutions to replace steady-state equations in COMP and AGG models.
Common lumped parameter models are reviewed in \cite{Shah.2003}.
{The work}~\cite{Luo.2002} 
presents an analytic solution to the stationary multi-stream HX equations under simplifying assumptions.
Further, HX networks may be used as a surrogate model of multi-stream HX \cite{Yee.1990,Hasan.2009}. 

\subsubsection{Reactors}
There are two main aspects that can result in high-order FOMs of chemical reactors. 
First, reaction systems can include high-dimensional multispecies kinetics with numerous reaction pathways \cite{Maas.1992,Vlachos.1996,Reizman.2012}.
Second, the spatial distribution of a reactor can prohibit lumped models, {\eg}, for non-ideally mixed tanks or for tubular reactors \cite{Shvartsman.1998,Hoo.2001}.

For kinetic systems, a reduction in the number of species corresponds to MOR, whereas a reduction in the number of modeled reactions may be regarded as hyperreduction.
Analogously to the simplifying assumptions discussed for column and HX modeling, the FOM of a kinetic system is often already a simplification of the real mechanisms and so is the mechanistic model of a reactor.
Naturally, for two-timescale kinetic systems with a clear partitioning between slow and fast dynamics, SPT can be directly applied \cite{Vora.2001}. 

A large set of works has been devoted to constructing invariant manifolds by solving the invariance equations of chemical kinetic systems with non-obvious slow and fast partitioning.
These methods employ system linearization \cite{Maas.1992},
computational singular perturbation \cite{Lam.1994},
thermodynamic dissipativity and symmetry \cite{Gorban.2004}
relation graphs \cite{Lu.2005,Pepiot.2008},
or sensitivity analysis and numerical optimization \cite{Edwards.1998,Petzold.1999,Bhattacharjee.2003}.
Extensive reviews are given in \cite{Gorban.2003,Chiavazzo.2007,Snowden.2017}.
Recently, {machine learning} {has} been increasingly applied to reduce kinetic systems \cite{Ji.2021,Wen.2023,Kasiraju.2023,Pateras.2025}.
{Interestingly, the chemical reaction network formalism has also been applied as mathematical framework to facilitate quadratization of generic ODEs \cite{Hemery.2020,Fages.2025}.}

For spatially distributed reaction systems, {\eg}, reaction diffusion problems and tube reactors,
the application of POD-Galerkin method 
\cite{Park.1996b,Krischer.1993,Hoo.2001,Armaou.2002}, 
nonlinear-Galerkin method \cite{Foias.1988b,Graham.1996,Shvartsman.1998}, manifold-Galerkin method \cite{Lee.2020},
and lift-and-reduce techniques \cite{Kramer.2019,Benner.2020}
have been reported.
As shown in \cite{Lee.2020}, nonlinear subspace methods can significantly outperform linear basis methods in the very-low-order region.
The classical collocation method is applied, {\eg}, in \cite{Hansen.1971,Georgakis.1977,Lefevre.2000}, 
to obtain low-order solutions of respective {partial differential equations}.
Moreover, an overview of compartment modeling of distributed chemical reactors is found in \cite{Haag.2018}.

For integrated reaction-separation systems (with recycle), the works \cite{Christofides.1996,Kumar.2002}
performed a time scale analysis by means of SPT to derive ROMs for control.
Along the same lines, \cite{Baldea.2006} investigated integrated reaction-{HX} networks.
Ref.~\cite{Schlegel.2002} compared the POD-Galerkin and POD-Res method on an integrated reaction-separation system, finding similar ROM accuracy but higher CPU costs of POD-Res ROMs.

\section{Case study}
\label{sec:case-study}
We compare the performance of MOR methods on an industrial process by studying the {air separation unit} presented in \cite{Caspari.2020}.
Clearly, comparing all the methods reviewed in Sections \ref{sec:methods} and \ref{sec:pse} would exceed the scope of this article.
Hence, we examine a subset of methods listed in Table \ref{tab:comparison}.
Specifically, we compare the MOR approaches (a-g) and (j). 
Additionally, we consider the two tailored methods COMP and AGG from Section \ref{sec:pse}.

We briefly comment on {this choice of methods}.
Despite successful applications in the literature, we investigate {neither} COL nor {nonlinear wave propagation models},
because {the former} are known to exhibit a significant steady-state offset \cite{Benallou.1986} and {the latter are} a special case of COMP.
We refer to \cite{Cao.2016,CaspariWave} for applications of {these methods} to the {process} considered herein.
{As discussed in Section \ref{sec:lift-and-learn}, we regard KW and DMDc as a variant of operator inference (h) and do not investigate other forms herein.}
Of the many possible approaches (i) and (j) reviewed in Section \ref{sec:ml-nonintrusive}, we choose the MFL-ANN method \cite{Masti.2021} due to conceptual simplicity and low implementation efforts.
Further, class (i) is theoretically included in (j) and therefore not investigated separately.
In particular, we expect all single-stage methods in Section \ref{sec:ml-nonintrusive} to reach a comparable accuracy.
Finally, recall that POD-Res and COMP represent data-driven and heuristic variants of SPT, respectively.
In this regard, our case study covers SPT.
As discussed in Section \ref{sec:intro}, we focus on MOR and do not investigate a potential hyperreduction and CPU times.

\begin{figure}[t!]
\centering
\includegraphics[width=0.7\linewidth]{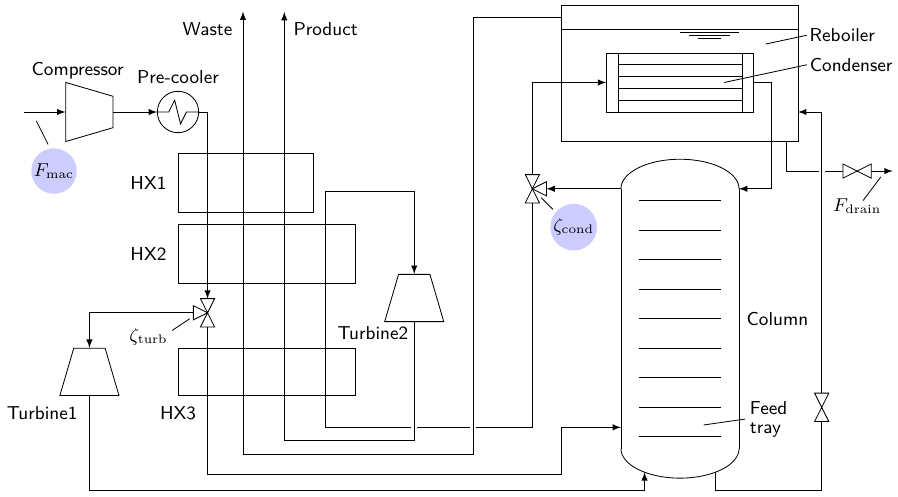}
\caption{{Schematic process flowsheet of the investigated air separation unit.}
Blue circles indicate {the system} inputs $\vec u(t)$.}
\label{fig:asu}
\end{figure}

Figure \ref{fig:asu} illustrates the {process} under investigation, which produces gaseous nitrogen and
is built of a main air compressor with precooling, 
three multi-stream HXs, two turbines, 
a high-pressure rectifying column with 30 trays,
and an integrated reboiler{-}condenser unit.
The {unit} has four manipulated {input} variables, 
which are the molar flow rate of air $F\ind{mac}$,
the molar flow fraction to the turbine $\zeta\ind{turb}$,
the reflux fraction at the column top $\zeta\ind{cond}$,
and the purge stream $F\ind{drain}$.

The {system} is open-loop unstable due to the integrator behavior of the liquid molar holdup $M\ind{irc}$ of the {reboiler tank}.
For unstable processes, data sampling is more challenging, especially if only a small subregion of the state space is relevant for ROM application.
However, most chemical plants feature a stabilizing base layer as part of the automation hierarchy, so stability is not an issue \cite{Engell.2012}.
For simplicity, we thus assume that the {reboiler tank} is equipped with an ideal inventory controller that manipulates $F\ind{drain}$ to stabilize $M\ind{irc}$ at its nominal value. 
Additionally, we fix $\zeta\ind{turb}$ to its nominal value.
Consequently, we have two variable inputs $\vec u = [F\ind{mac}, \zeta\ind{cond}]\transp$.

For our purpose, the selected {process} features some characteristic properties of industrial process systems, such as variable coupling and nonlinear process response. 
Moreover, we regard the {process} as a good compromise between the complexity of the process system and the complexity of the MOR problem.
However, the selected process does not incorporate a chemical reactor and has few inputs.
The investigation of MOR for complex processes including reaction and many {input variables} is indispensable and should be undertaken in future studies.

\subsection{Full-order model}
We consider the three main species of air nitrogen, argon, and oxygen.
In contrast, we do not model carbon dioxide, hydrocarbons, or other residual chemical species here.
The FOM comprises material and energy balances of all process units as well as constitutive equilibrium-based thermodynamic equations.
The {column} is modeled in a tray-to-tray fashion assuming thermal equilibrium with constant relative volatilities,
constant specific heat capacities of all species, 
and ideal mixture (no excess properties).
Both the {column} bottom tray and the {reboiler}
are modeled as special column trays with an extra feed stream and fixed liquid outflow, respectively.
The condenser is a total condenser with negligible material holdup and modeled using lumped quasi-stationary balance equations.
We apply a logarithmic transformation of molar fractions to support the extraction of composition patterns, {\ie}, the logarithmic molar fractions are the respective differential states. 
The HX models feature energy balances for the metal wall and quasi-stationary balances for the fluids, assuming a single {flow chanel for each stream}, 1D spatial discretization with $n\ind{HX}$ cells per stream, and the common wall assumption.
Following \cite{Caspari.2020}, we set $n\ind{HX1}= 25$, $n\ind{HX2}= 25$, and $n\ind{HX3}= 1$.
Lastly, first-order low-pass filters account for the lagged vapor flow rate in {heat exchangers} and {column}.

In total, the FOM comprises $n_u=2$ inputs and $n_x=176$ differential states, and is implemented in \texttt{Modelica}.
{Although $n_x$ may appear low compared to some structural mechanic or fluid dynamics problems, the system investigated here constitutes a representative example of model reduction in the process control field, \cf~\cite{Schulze.2021,Quarshie.2023,ElWajeh.2024}.}
The admissible input set $\set U$ is a box defined by $F\ind{mac} \in [\unit[30]{mol/s},\unit[50]{mol/s}]$ and $\zeta\ind{cond}\in[0.51, 0.54]$.
As the initial state $\vec x_0$,
we use the nominal steady state corresponding to $F\ind{mac}=\unit[52.2]{mol/s}$, $\zeta\ind{cond}=0.535$, $\zeta\ind{turb}=0.805$, and $M\ind{irc} = \unit[25]{kmol}$.
At $\vec x_0$, the {process} produces \unit[25]{mol/s} nitrogen of \unit[100]{ppm} impurity grade.

\subsection{Data sampling}
In order to guarantee a fair comparison of snapshot-based methods against each other,
we always use the same training data.
This data set is recorded by simulating the FOM subject to randomly ordered inputs drawn from an equidistant $5 \times 5$ grid in $\set U$.
Following \cite{Caspari.2020}, the sampling time is $\Delta t\ind{s} = \unit[15]{min}$.
We impose the input sequence as steps (zero-order hold) of \unit[240]{h} (10 days) {each}.
Thereby, we include information from both the fast spectrum of the system response and the stationary states.
We illustrate the input profile $\vec u(t)$ in the supplementary information.

We simulate the FOM in \texttt{Dymola 2025x} using the integrator \texttt{DASSL} \cite{Petzold.1982} with integration tolerance $10^{-6}$.
The recorded data set comprises $N=24\,000$ snapshots of each state.
The corresponding snapshot matrices have the form 
$\mat X \in \R^{n_x \times N}$
and $\mat U \in \R^{n_u \times N}$.
These snapshots cover the operating space between \unit[15]{mol/s} and \unit[27]{mol/s} production rate and product quality between \unit[10]{ppm} and \unit[1000]{ppm} impurities.

\subsection{Implementation}
As previously discussed, POD and MFL are strongly affected by scaling.
Herein, we scale all states and inputs individually to the range $[-1,1]$ and apply zero-mean centering.
We remark that a tailored scaling strategy, {\eg}, scaling based on physical relevance and range, may enable more accurate ROMs (with respect to a target application).
Like the FOM, we implement all intrusive ROMs in \texttt{Modelica} and perform all simulations in \texttt{Dymola} using \texttt{DASSL} with an integration tolerance of $10^{-6}$.
The non-intrusive ROMs (DMDc, KW, MFL-ANN) are directly evaluated in the respective {machine learning} environments in \texttt{Python} (see below).
Next, we provide further details on the implementation.

\subsubsection{POD methods and DMDc}
We implement POD and DMDc using \texttt{SciPy} as a \texttt{Python} interface to the SVD implementation in \texttt{LAPACK}. 
All other matrix computations are performed using \texttt{Numpy}.
We specify \texttt{float64} precision, which was found to be crucial for obtaining reliable SVD results. 
We scale the latent states $\vec z(t)$ heuristically, which was found to improve model conditioning.
For DMDc, we calculate the coefficient matrices $\mat A$, $\mat B$, and $\mat U_1$ in \eqref{eq:dmdc-rom} using the formulas given in \cite{Proctor.2016}.
To derive a POD-Res model, we use a truncated residualization, where $n_q = 50-n_z$.
The truncation improves simulation convergence without significant loss of precision.

\subsubsection{MFL-Galerkin and MFLu-Galerkin methods}
\label{sec:mfl-implementation}
We train AENs in \texttt{Tensorflow} using the standard training problem \cite{Vincent.2008}.
To reduce the number of trainable parameters and speed up the training, we prereduce the scaled and zero-centered snapshots by projecting the data onto the first $100$ POD modes \cite{Phillips.2021}.
For the success of MFL, it is crucial to not rescale the {POD-}projected snapshots again.
The encoders and decoders have a single fully connected nonlinear hidden layer with 50 neurons with (smooth) \texttt{tanh} activation, respectively.
Using deeper networks or additional neurons did not improve the reconstruction accuracy considerably.

We divide the pre-processed snapshots into \unit[80]{\%} training data and \unit[20]{\%} validation data.
For every network, we run three trainings with random seeds and select the weights with the best performance on the validation data set.
We train all networks for a total of $20\,000$ epochs and minibatch size 256, using \texttt{Adam} with initial learning rate of $10^{-4}$ for the first $15\,000$ epochs and \texttt{SGD} with {learning rate} of $10^{-6}$ for subsequent finetuning.
For the implementation of MFLu, we modify the AEN structure from $(\vec \varphi, \vec \varphi^\dagger)$ to $(\vec \psi, \vec \psi^\dagger)$ as discussed in Section \ref{sec:ml-control}.
However, the general training setup remains the same.
To include the AENs in the ROMs, we implement the decoder network structure in \texttt{Modelica}.

\subsubsection{KW models}
\label{eq:kw-training}
We use our {machine learning} framework \cite{Schulze.2022a} to train reduced KW models.
This framework is based on \texttt{Tensorflow} and implements the KW model structure along with a training procedure based on snapshot data.
As in Section \ref{sec:mfl-implementation}, we pre-reduce the training data through projection onto the principal $100$ POD modes.
The trajectory snippets required for the prediction loss are created by sliding along the training data set in a moving horizon fashion, stopping every 10 samples and copying a series of 100 consecutive snapshots.
The encoder and decoder networks have a symmetric structure with two hidden layers and linear output layer.
The number of hidden neurons are determined via linear interpolation between $100$ and $n_z$.
We use \texttt{ELU} activation and train the model for $10\,000$ epochs and minibatch size 32, where the first $8\,000$ epochs use \texttt{ADAM} with {learning rate} of $10^{-3}$ and the final $2\,000$ epochs use \texttt{SGD} with {learning rate} of $10^{-6}$ for finetuning.
All other hyperparameters have the default values from
\cite{Schulze.2022a}.
The ROM simulation employs the graph-based execution mode of \texttt{Tensorflow}.

\subsubsection{MFL-ANN models}
We implement the method in \cite{Masti.2021} by means of a modification of our aforementioned framework \cite{Schulze.2022a}.
To this end, we substitute the linear dynamics in \eqref{eq:KW} by a feedforward ANN to learn the discrete-time latent predictor:
\begin{equation}
    \vec z_{k+1} = \vec F(\vec z_k, \vec u_k),
\end{equation}
where $\vec F: \set Z \times \set U \rightarrow \set Z$ and $k\in\{0,1,...\}$.
The ANN $\vec F$ features two hidden layers of each $2n_z$ neurons and \texttt{ReLU} activation, and a linear output layer.
This specification is parsimonious compared to \cite{Masti.2021}, who use three hidden layers of each $5n_z$ neurons.
However, we found that overparameterization results in poor training convergence.
All other settings, including AEN structure and training hyperparameters, are identical to \eqref{eq:kw-training}.

\subsubsection{COMP and AGG models}
Both methods are applied to the {FOM} in a unit-by-unit fashion.
However, the submodels of HX3, {column} feed stage, and {reboiler/condenser} already have a minimal form and cannot be further reduced by the COMP or AGG.
Since the dynamic relevance of these units is unclear without further investigation, we leave them untouched.
Moreover, preliminary experimentation showed that the HX dynamics have a noticeable effect on the overall system response.
Hence, we do not replace these units by quasi-stationary models, {\ie}, we create and insert ROM submodels for HX1, HX2, and {column}.

The number and size of compartments and the position of the sensitive elements (referring to {column} trays and HX cells simultaneously)
are degrees of freedom of the COMP method and may be regarded as hyperparameters.
Similarly, the hyperparameters of AGG are the number and positions of aggregation elements.
We follow \cite{Schafer.2019} and use compartments of a uniform size, with sensitive elements located in the center of the respective compartments.
A perturbation analysis of the positions confirmed this choice.
We select the same central positions for the aggregation elements. 
Using the same number of compartments or aggregation elements $n\ind{c}$ for each HX1, HX2, and {column}, 
we obtain $n_z = 10 + 6\cdot n\ind{c}$.
In all AGG models, we specify $\mat H$ in \eqref{eqn:aggr-rom} such that $\mathrm{diag}(\mat H) = n_x$.

\subsection{Definition of the comparison criteria}
We compare the selected MOR techniques in two tests.
First, we only assess the accuracy of the linear or nonlinear subspaces, {\ie}, the slow manifold approximations,
constructed by the MOR methods.
This assessment does not yet include ROMs but merely determines the average accuracy of projecting the FOM state snapshots of a test data set onto the respective low-dimensional subspaces.
In particular, the projection error constitutes a lower bound on the ROM prediction error \cite{Rathinam.2003}.
Second, we simulate the ROMs, {\ie}, we examine the prediction capabilities of the ROMs.
This prediction test is the most important criterion as we assess the overall ROM accuracy.
We initialize all ROMs at the nominal steady state described above.

In both evaluations, we use an independent test data set that comprises a series of random uniformly distributed zero-order hold inputs (illustration see {supplementary information}).
Furthermore, we consider different degrees of reduction, {\ie}, we vary the reduced order $n_z$.
As previously discussed, for COMP and AGG,
the values of $n_z$ are not freely selectable but are restricted by the subsystem-by-subsystem application of the methods, where viable options are $n_z\in \{16, 22, 28,...\}$.
To facilitate a comparison of the MOR methods, we include such values of $n_z$ into the study.

We evaluate the accuracy in terms of the root mean squared error \abk{RMSE},
which we define in terms of the Frobenius norm:
\begin{equation}
    \mathrm{RMSE}(\mat X, \mat X\ind{r}) \defn \sqrt{\frac{1}{n_xN}\lVert \mat X - \mat X\ind{r} \rVert\ind{F}^2} .
\end{equation}
Therein, $\mat X\in \R^{n_x \times N}$ is the scaled test snapshot matrix, having the snapshots $\vec x^{(k)}$, $k=1,...,N$, as matrix rows.
The matrix $\mat X\ind{r}\in \R^{n_x \times N}$ collects the projected or predicted snapshots, $\vec x\ind{r}^{(k)}\in \R^{n_x}$, $k=1,...,N$.
We scale and shift $\mat X$ and $\mat X\ind{r}$ consistently with the training data.

We note, however, that the choice of the error norm and test data set can have a significant impact on the conclusions and should always be selected based on the ROM application.
For example, the RMSE averages out peak deviations so that the slow and steady-state errors are prioritized over maximum errors, {\eg}, an {inverse response} on a short-term scale does not drastically increase the error.
Herein, our test data set comprises a significant share of slow response and near-stationary snapshots.
We consider this choice to be suitable for examining ROMs that capture the slow manifold.

Finally, we describe how the projections for test 1 are obtained.
Regarding the linear basis methods POD and DMDc, both the linear subspaces and the projections coincide, respectively. 
The projection is given by the projection matrix, $\mat \Pi \defn\mat U_1 \mat U_1\transp $, where $\mat U_1$ is from \eqref{eqn:pod-expansion} or \eqref{eq:dmdc-rom}.
To compute the projections performed by MFL, MFLu, KW, and MFL-ANN, we encode and decode the test data through the respective AENs.
The projections in POD-Res, COMP and AGG are inherently defined through the degeneration/residualization.
We evaluate these projections by solving the respective nonlinear systems of equations, Eqs.~\eqref{eqn:pod-residual-stationary}, \eqref{eqn:spt-ae} and \eqref{eq:agg-ae},
on the test data in \texttt{Dymola}.
Recall that for the selected hyperparameters, the slow manifolds of AGG and COMP coincide; however, the projections differ.

\subsection{Results: Projection accuracy}
We present the results of the projection test.
Table \ref{tab:projection} compares the projection RMSE on the test data set.
First, we consider small $n_z\leq10$, {\ie}, the very-low-order region.
As expected, the nonlinear subspace methods (POD-Res, MFL, MFLu) perform significantly better than the linear basis methods (POD, DMDc).
The respective difference in RMSE is approximately one order of magnitude.
Overall, the smallest projection errors are reached by the MFLu approach.
However, the difference in RMSE between MFL and MFLu decreases with increasing $n_z$.
This confirms {the input-dependency of the slow manifold as discussed in} Section \ref{sec:ml-control}.
Clearly, the error difference is system-dependent and may be even more substantial for a strongly parameterized slow manifold.
Finally, the projection RMSEs of the non-intrusive nonlinear subspace methods KW and MFL-ANN {are} comparable to POD.
The higher RMSE than for MFL is attributed to the additional loss terms included in the KW and MFL-ANN trainings. 

\begin{table}[t!]
    \centering
    \caption{Results of projection accuracy. 
    The RMSE indicates the average error between the states projected on the reduced subspace versus the original FOM states. 
    We consider different ROM orders $n_z$.}
    \label{tab:projection}
    \resizebox{\textwidth}{!}{
    \begin{tabular}{lcccccccc}
    \toprule
    $\mathbf{n_z}$ & \textbf{POD/DMDc} & \textbf{POD-Res} & 
    \textbf{MFL} & \textbf{MFLu}  & 
    \textbf{KW} & \textbf{MFL-ANN} & 
    \textbf{COMP} & \textbf{AGG}\\
    \midrule
    2  & $7.8 \cdot 10^{-2}$ & 
    $4.9 \cdot 10^{-2}$ &
    $1.0 \cdot 10^{-2}$ & % MFL
    $5.3 \cdot 10^{-3}$ &  % MFLu
    $ 1.5 \cdot 10^{-1}$ & % KW
    $ 1.6 \cdot 10^{-1}$ & % MFL-ANN
    $-$ & $-$
    \\
    5  & $1.6 \cdot 10^{-2}$ & 
    $1.7 \cdot 10^{-2}$ &
    $2.2 \cdot 10^{-3}$ & % MFL
    $1.8 \cdot 10^{-3}$ & % MFLu
    $1.4 \cdot 10^{-2}$ & % KW
    $2.5 \cdot 10^{-2}$ & % MFL-ANN
    $-$ & $-$
    \\
    10 & $3.8 \cdot 10^{-3}$ & 
    $1.2 \cdot 10^{-2}$ &
    $1.3 \cdot 10^{-3}$ & % MFL
    $1.4 \cdot 10^{-3}$ & % MFLu
    $7.3 \cdot 10^{-3}$ & % KW
    $2.5 \cdot 10^{-2}$ & % MFL-ANN
    $-$ & $-$
    \\
    16 & $1.1\cdot 10^{-3}$ & 
    $ 5.5 \cdot 10^{-3}$ &
    $7.3 \cdot 10^{-4}$ & % MFL
    $1.5 \cdot 10^{-3}$ & % MFLu
    $6.1 \cdot 10^{-3}$ & % KW
    $1.5 \cdot 10^{-2}$ & % MFL-ANN
    $9.4 \cdot 10^{-2}$ & % COMP
    $4.8 \cdot 10^{-2}$ % AGG
    \\
    28 & $9.0 \cdot 10^{-5}$ & 
    $3.6 \cdot 10^{-3}$ & 
    $1.4 \cdot 10^{-3}$ & % MFL
    $1.5 \cdot 10^{-3}$ & 
    $5.9 \cdot 10^{-3}$ & % KW
    $1.5 \cdot 10^{-2}$ & % MFL-ANN
    $1.8 \cdot 10^{-2}$ & % COMP
    $2.0 \cdot 10^{-2}$ % AGG
    \\
    \bottomrule
    \end{tabular}
    }
\end{table}

Moving to higher orders, {\ie}, $n_z\in\{16, 28\}$, 
the improvement by the MFL-based methods is only moderate, which was also observed in \cite{Lee.2020,Phillips.2021}.
We attribute this limited improvement to the challenging convergence of the respective unsupervised learning problems using local solvers, {\ie}, training AENs by stochastic gradient descent algorithms.
As a result, POD truncation outperforms the MFL methods at $n_z=28$, facilitating the overall smallest projection error.
Interestingly, the linear POD subspace projection also exceeds the accuracy of the nonlinear POD-Res method.
A further investigation of the POD-Res results revealed {notable} deviations between $\vec x(t)$ and $\vec x\ind{r}(t)$ at the points of abrupt changes in $\vec u(t)$. 
Finally, both COMP and AGG exhibit relatively large errors compared to the other ROM methods.
Here, the COMP subspace shows significant deviations from the FOM samples near the input steps but relatively small errors otherwise.
In contrast, the maximum deviations of AGG are lower, but a greater number of samples have a notable projection error.

\subsection{Results: ROM predictions}
We now present the results of the ROM prediction test.
Table \ref{tab:results-1} collects the state prediction RMSE of the ROMs.
The RMSE {in Table \ref{tab:results-1}} is consistently higher than in Table \ref{tab:projection},
because ROM predictions are less accurate than the {mere} projection of FOM snapshots onto the ROM subspace.

Both linear subspace methods, POD-Galerkin and DMDc, 
do not reach the desired accuracy at very low order, {\ie}, $n_z\leq10$.
For reduced orders $11\leq n_z \leq 21$, the simulations of the POD-Galerkin ROMs did not converge at all.
As we discuss in the {supplementary information}, these critical values of $n_z$ lie below the truncation threshold of $n_z= 23$, obtained from the POD singular values.
At order $n_z=28$, the POD-Galerkin ROM successfully provides an accurate prediction, whereas the DMDc prediction is still poor.
Clearly, the nonlinearity of the FOM prohibits highly accurate DMDc models.
Finally, we notice a significant gap between projection error (Table \ref{tab:projection}) and projection error (Table \ref{tab:results-1}) for both POD-Galerkin and DMDc.

Compared to the POD-Galerkin results, the POD-Res model is more accurate throughout all orders.
This advantage is particularly significant in the very-low-order region.
In contrast to the projection accuracy (Table \ref{tab:projection}), POD-Res outperforms the POD-Galerkin ROMs in the prediction.
Even for $n_z=2$, we notice a fairly small prediction error of POD-Res.
Since an RMSE of $10^{-2}$ roughly corresponds to \unit[1]{\%} average error in the scaled states, we expect a satisfactory {precision} of POD-Res ROMs of order $n_z=10$ (and possibly even $n_z=5$) in {tasks such as control}.
When moving to higher $n_z$, the accuracy improves further.

\begin{table}[t!]
    \centering
    \caption{Results of ROM predictions. 
    The RMSE indicates the average error between the state predictions by the ROMs versus the true FOM states.
    We consider different ROM orders $n_z$.
    Non-converged simulations are indicated by $(*)$.}
    \label{tab:results-1}
    \resizebox{\textwidth}{!}{
    \begin{tabular}{lccccccccc}
    \toprule
    $\mathbf{n_z}$ & \textbf{POD} & \textbf{DMDc} & \textbf{POD-Res} &
    \textbf{MFL} & \textbf{MFLu}  & \textbf{KW} & \textbf{MFL-ANN} & \textbf{COMP} & \textbf{AGG}\\
    \midrule
    2 & $1.1 \cdot 10^{-1}$ & 
    $1.5 \cdot 10^{-1}$ & % DMDc
    $3.6\cdot 10^{-2}$&
    $6.1\cdot 10^{-1}$ & % MFL
    $2.9\cdot 10^{-1}$ & % MFLu
    $1.5\cdot 10^{-1}$ & $ 1.7 \cdot 10^{-1}$ &
    $-$ & $-$
    \\
    5 & $ 6.3 \cdot 10^{-2}$ 
    & $1.5 \cdot 10^{-1}$ & % DMDc
    $2.1\cdot 10^{-2}$ &
    $1.4\cdot 10^{-1}$ & % MFL
    $ 2.2\cdot 10^{-2}$ & % MFLu
    $ 1.5\cdot 10^{-2}$ &  % KW
    $ 2.9 \cdot 10^{-2}$ &  % MFL+ANN
    $-$ & $-$
    \\
    10 & $4.9\cdot 10^{-2}$ & 
    $1.3 \cdot 10^{-1}$ & % DMDc
    $8.8 \cdot 10^{-3}$ &
    $2.1 \cdot 10^{-2}$ & % MFL
    $1.5\cdot 10^{-2}$ & % MFLu
    $9.9 \cdot 10^{-3}$ & % KW
    $3.3 \cdot 10^{-2}$ & % MFL-ANN
    $-$ & $-$
    \\
    16 & $(*)$ & 
    $9.4 \cdot 10^{-2}$ & % DMDc
    $7.9 \cdot 10^{-3}$& % POD-Res
    $1.0\cdot 10^{-2}$ & % MFL
    $ 1.1\cdot 10^{-2}$& % MFLu
    $1.9 \cdot 10^{-2}$ & % KW
    $3.2 \cdot 10^{-2}$ &  % MFL-ANN
    $2.5 \cdot 10^{-1}$ & % COMP
    $4.9 \cdot 10^{-2}$ % AGG
    \\
    28 & $8.7\cdot 10^{-3}$ & 
    $7.8 \cdot 10^{-2}$ & % DMDc
    $7.5 \cdot 10^{-3}$ &
    $(*)$ & % MFL
    $(*)$ & % MFLu
    $1.1 \cdot 10^{-2}$ & % KW
    $3.3 \cdot 10^{-2}$ & % MFL-ANN
    $3.7 \cdot 10^{-2}$ & % COMP
    $1.7 \cdot 10^{-2}$ % AGG
    \\
    \bottomrule
    \end{tabular}
    }
\end{table}

Similarly to POD-Res, the methods MFL and MFLu enable a higher precision than the linear basis methods at very low order (except for $n_z=2$).
Moreover, at $n_z=5$, the MFLu-Galerkin ROM clearly outperforms the MFL-Galerkin ROM.
On the other hand, at higher order, both models perform similarly.
At $n_z=28$, the simulation of both MFL and MFLu ROMs did not converge.
Using a different activation function or adding neurons to the AENs did not resolve this issue.
A possible cause is the combination of relatively high ROM order at moderate AEN accuracy (similarly to the issues with POD-Galerkin ROMs).
Although we are confident that modifications based on more extensive training might facilitate convergence, we did not undertake such attempts.
Instead, we highlight the potential numerical issues with higher-order MFL-Galerkin ROMs when working in a standard AEN training setup.

The non-intrusive KW and MFL-ANN ROMs have a prediction error comparable to the intrusive nonlinear subspace methods.
{For each $n_z$,} the differences in RMSE {are insignificant, where KW models perform only slightly better}.
Notably, the MFL-ANN models do not perform better than the KW models.
A comparison to the performance on the training data set revealed a tendency of MFL-ANN to overfit the training data (the RMSEs on the training data were approximately one order of magnitude smaller than on the test data), which was less pronounced for the KW structure.
From $n_z=10$ to $n_z=28$, there is no improvement in RMSE for both model types, {\ie}, increasing the ROM order does not improve the accuracy.
Finally, we notice that the projection and prediction errors are comparable for both models, respectively.

For COMP and AGG models with a single compartment and aggregation cell for each unit operation, respectively, we find significant prediction errors.
In particular, the COMP model is far from the desired accuracy due to severe {inverse response}.
Increasing the ROM order results in a notably smaller RMSE, where the AGG results reach the desired error level.
However, the COMP response still exhibits a notable {inverse response}, diminishing ROM accuracy (see {Fig.~\ref{fig:results}} below).

Besides comparing the numerical results, we also comment on the numerical properties of the ROMs.
Performing the simulations was most straightforward using the non-intrusive ROMs (DMDc, KW, MFL-ANN), which are numerically favorable to deal with due to the sequential model structure.
On the other hand, simulating the POD-Res and COMP models was computationally most expensive, and required a very careful initialization and scaling of the model equations to converge the simulations.
As a result, the high accuracy of the POD-Res model is offset by the fact that it is difficult to handle numerically.

\begin{figure}[ht!]
	\centering
    \begin{subfigure}{0.5\linewidth}
    \includegraphics[width=\linewidth]{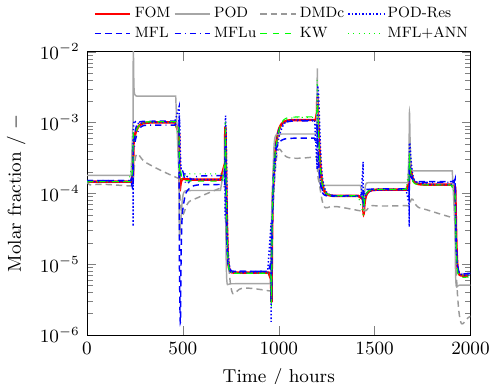}
    \caption{}
    \label{fig:results-a}
	\end{subfigure}
	\hfill
    \hspace{-10pt}
    \begin{subfigure}{0.5\linewidth}
    \includegraphics[width=\linewidth]{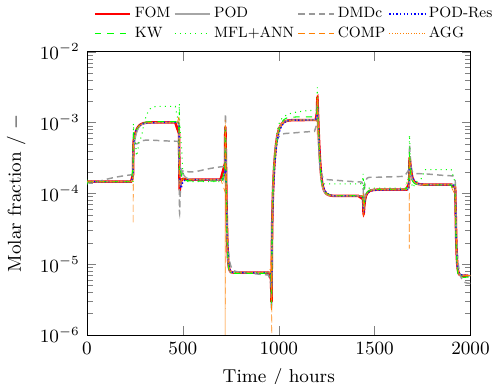}
    \caption{}
    \label{fig:results-b}
	\end{subfigure}
	\caption{Results of test 2 (prediction). 
    Depicted is the liquid molar fraction of oxygen on the {column} top stage over the first \unit[2000]{h} of the test.
    ROM order is (a) $n_z = 5$ and (b) $n_z=28$.
    }
	\label{fig:results}
\end{figure}

To illustrate the results, we examine the molar fraction of oxygen on the {column} top tray, which is directly connected to the quality of the nitrogen product.
We show the predictions for $n_z=5$ in Fig.~\ref{fig:results-a} and for $n_z=28$ in Fig.~\ref{fig:results-b}.
In Fig.~\ref{fig:results-a}, we observe a significant offset between ROM prediction and FOM response for POD, DMDc, and MFL.
This finding is consistent with Table \ref{tab:results-1}.
For these ROMs, the offset involves both a transient and a steady-state mismatch.
Moreover, despite the relatively small RMSE of POD-Res, the graph shows significant {inverse response} near the input steps, whereas the steady states are captured accurately.
All other ROMs (MFLu, KW, MFL-ANN) reproduce the FOM trajectory closely.

Observing the higher-order case in Fig.~\ref{fig:results-b},
we find significantly better predictions by the POD and POD-Res models.
At the same time, the DMDc forecast remains inaccurate.
The effect of overfitting discussed above for the MFL-ANN ROMs is also visible.
Compared to Fig.~\ref{fig:results-a}, the {MFL-ANN} prediction is less accurate despite a higher-order model.
Further, notable deviations from the FOM trajectory are visible for the COMP model in the form of severe {inverse response}.
In contrast, the AGG ROM provides a close reproduction of the FOM trajectory.

\section{Discussion}
\label{sec:case-study:comparison}
Based on theoretical properties of the MOR methods and the findings of the case study, we summarize characteristic properties of the methods investigated.
However, we emphasize that definitive statements on method ranking are delicate, because in addition to $n_z$, most MOR techniques possess further hyperparameters that can strongly affect ROM behavior, {\eg}, the position of the sensitive element (COMP, AGG) or the ANN architecture (MFL, KW, MFL-ANN).
Although we performed a basic validation of crucial hyperparameters within the case study, we do not claim that this screening was exhaustive.
Additionally, the performance of a certain type of ROM may be system-dependent, and in that sense our experimental results may be specific to air separation {units}.

Comparing the results of intrusive (MFL, MFLu) to non-intrusive nonlinear subspace methods (KW, MFL-ANN) in the case study, 
we conclude that similar ROM {accuracy is} reached when targeting very low-order models.
Such very-low-order ROMs are important in control applications, where CPU reduction has a high priority whereas moderate errors are acceptable.
Often, a moderate prediction error can be compensated by state estimation and model adaptation, especially if the error is predominantly steady-state offset \cite{Morari.2012}.
On the other hand, despite a relatively low RMSE, the POD-Res results exhibited an {inverse response} in the very-low-order setup, which can be problematic in control applications.

When prioritizing accuracy over low order, non-intrusive methods may {face difficulties} to reach arbitrary precision due to their inherently difficult learning problems. 
In such cases, intrusive strategies can be {easier} to apply.
In particular, the POD-based methods were able to generate very accurate ROMs given sufficiently high order.
In this context, we acknowledge the simplicity and approximation guarantees (on the training data) of the POD.
In contrast, high projection and prediction accuracies were difficult to reach when using MFL strategies.

The advantage of non-intrusive over intrusive reduction methods is that non-intrusive strategies combine MOR and hyperreduction.
The weakness of non-intrusive methods is that they can be prone to overfitting if not carefully regularized.
One possible \hl{structural regularization} strategy is to restrict the latent dynamics to a linear or bilinear form.
In particular, applied Koopman theory combines MFL and linearization into a KW structure, which enables reliable ROMs of the moderately nonlinear systems.

POD-Res and COMP may be considered as data-driven and heuristic variants of SPT, respectively.
However, COMP retains most of the model sparsity, whereas POD-Res ROMs are generally dense and can therefore be more challenging to solve.
Here, both POD-Res and COMP were numerically the most delicate to handle.
Specifically, large-scale implicit nonlinear systems of equations and non-physical {inverse response} made these models difficult to converge.

In terms of reduction effort, the (non-intrusive) KW and MFL-ANN strategies involved the most extensive offline computations.
Achieving {sufficiently} accurate models may require multiple trainings and extensive hyperparameter tuning, which introduces delays into the model reduction procedure.
Furthermore, non-intrusive methods are most sensible to the training data in our experience. 
In particular, a balanced amount of transient and stationary data is essential for the successful generation of ROMs that enable precise long-term forecasts.

The model reduction efforts in intrusive reduction methods (including tailored methods) demand a great share of model building and reformulation (debugging, robustification). 
On the other hand, the {training} problems used within intrusive methods, {\eg}, AEN training, are typically less computationally demanding than in non-intrusive approaches.
In our experience, intrusive data-driven methods are slightly more tolerant to a poorly designed data set {(\eg}, too few steady-state data{) than non-intrusive methods}.
For intrusive methods, further efforts have to be expected in terms of hyperreduction, often necessary to obtain a real-time capable ROM. 

Lastly, we comment on the model reduction procedure for interconnected systems.
Applying the tailored methods, such as COMP and AGG, is naturally done in a \hl{subsystem-by-subsystem} fashion, since these methods provide reduced models of the subsystems, {\eg}, distillation columns or HXs, of an interconnected system, {\eg}, a process flowsheet model.
In particular, when working in a structured modeling environment,
such as \texttt{Unisim Design} or \texttt{gPROMS},
these ROMs can be pre-configured and included into the model library.
Reducing an interconnected system model is then achieved by substituting full-order submodels by reduced submodels.
The ``subsystem-by-subsystem'' approach preserves the model sparsity originating from the interconnection.
{However,} the feasible degree of reduction may be limited, as shown by the case study results.
Further, the reduction problem becomes a combinatorial problem due to potentially many subsystem ROMs to be configured and combined.

{From a mathematical perspective, the \hl{subsystem-by-subsystem} strategy falls into the port-Hamiltonian formalism.
Notably, in this framework, the respective ROMs are invariant under couplings and preserve their properties.
Hence, this formalism should be investigated in future work to develop more accurate alternatives to COMP and AGG.}

In contrast to the \hl{subsystem-by-subsystem} approach, 
general-purpose methods are most easily applied to the entire FOM.
Then, the complete model of the interconnected system is reduced as a whole.
As shown by the case study, such an approach can enable a lower ROM order.
However, for systems with many inputs, the data sampling efforts grow exponentially with $n_u$, which can result in a curse of dimensionality.
This curse may be broken by a \hl{subsystem-by-subsystem} strategy if the high-dimensional inputs enter the system only locally and are evenly distributed over the submodels.

\section{Conclusions and outlook}
\label{sec:conclusions}
Model order reduction spans 
a collection of techniques to construct 
a low-order model approximating the evolution of a high-order system.
Herein, we have reviewed and compared reduction methods for nonlinear dynamical models with inputs.
In particular, we have contrasted the different strategies both theoretically and in a case study.
As expected, there is no universal best choice when selecting a reduction method.
Rather, a suitable choice depends on targeted ROM properties.

If very-low-order ROMs are needed, then intrusive nonlinear subspace methods, {\eg}, the MFLu approach, 
as well as non-intrusive nonlinear subspace methods, {\eg}, the KW model,
have provided accurate ROMs within our case study.
Notably, non-intrusive reduction methods combine MOR and hyperreduction in a single step, providing a ROM that is real-time capable.
If high ROM accuracy is needed but the degree of reduction is of secondary importance, then linear subspace methods, {\eg}, the POD-Galerkin method, are a good choice due to their (theoretically) arbitrary degree of accuracy and simplicity of application.
Finally, when desiring structural preservation of a given network or flowsheet model comprising interconnected subsystems, 
then {structure-preserving (especially port-Hamiltonian) ROMs or} problem-specific ROMs are a straightforward choice{. These approaches} reduce the subunits individually by means of pre-configured ROM blocks.

Besides reviewing and comparing existing MOR methods, we have also proposed an extension of the manifold-Galerkin method to systems with inputs (Section \ref{sec:ml-control}).
In particular, we stress that system inputs $\vec u(t)$ can deform the slow manifold $\set M_{\set X}$.
Consequently, if the effect of inputs is disregarded, an accurate approximation of $\set M_{\set X}$ may necessitate a high ROM order.

Within the case study, we validated {the proposed manifold-Galerkin extension for systems with inputs. In cases of very strong reduction, we} found that our approach outperforms the {standard} method both in terms of the projection accuracy and the ROM prediction accuracy.
As expected, our extended approach and the established approach performed similarly at higher orders.

In our case study, we did not investigate hyperreduction or CPU cost of integration.
Instead, we expect that a sufficiently low order combined with hyperreduction enables real-time applications, {\eg},  \cite{Schafer.2019,Schulze.2023}.
Clearly, real-time capability depends on the application, so there is no universal CPU threshold to classify performance.
In any case, accounting for hyperreduction already within MOR could be a promising step towards more holistic methods.
{Even though our case study represents a realistic MOR problem in the process systems control field, future work should investigate higher-dimensional problems and examine additional reduction methods, \eg, \cite{Bychkov.2024,Geelen.2024}.}

Despite increasing computer power, nonlinear MOR (especially for parametric and non-autonomous systems) remains a challenging discipline worth of investigation.
The curse of dimensionality, when sampling systems with many inputs or parameters, is connected to the need for robust methods applicable in the low-data limit.
In this regard, {structure-preserving MOR is}
a promising approach to enrich the non-intrusive manifold learning task by additional ({physics-based}) information.
Moreover, {the scalability of} manifold learning methods to very-high-order problems remains an open challenge.
{Recently, promising results have been presented using quadratic/polynomial manifold learning \cite{Szalai.2023,Geelen.2024} and convolutional autoencoders \cite{Fresca.2022,Romor.2023}.}

{Although several methods offer stability preservation, such guarantees are not inherently provided by every method} \cite{Hesthaven.2022}.
Consequently, {a careful method selection or}
a-posteriori error and stability analysis are required.
{Another important topic is the development of a-priori error bounds on ROM accuracy.}

Finally, we notice some limitations of current reduction methods with respect to numerical optimization and control applications.
First, most numerical optimization techniques are based on derivative information, in particular, the Jacobian of the ROM variables $\vec z(t)$ and $\vec x\ind{r}(t)$ in \eqref{eq:rom} with respect to each other and the inputs $\vec u(t)$.
However, standard MOR methods do not explicitly account for the accuracy of such derivatives.
For learning-based MOR and hyperreduction approaches, Sobolev trainings \cite{Czarnecki.2017,Tsay.2021b} may improve derivative accuracy.
Additionally, in our experience, sensitivity integration is often considerably more expensive than standard integration, and therefore accounting for sensitivities in the reduction process might enable fundamental improvements in CPU costs of optimization applications.
Second, model reduction of non-smooth or even discontinuous models, {\eg}, for controlling a process start-up, {is a relevant} but mostly unexplored topic \cite{Yamaleev.2013,Bettini.2024}.

\section*{Acknowledgments}
A pre-print version of this manuscript has been posted on \url{https://arxiv.org/pdf/2506.12819} under the arXiv.org non-exclusive distribution license.
{The authors gratefully acknowledge the financial support of the Kopernikus project SynErgie by the German
Federal Ministry of Research, Technology and Space (BMFTR) and project supervision by Projekttr\"ager J\"ulich (PtJ).}
The authors thank Danimir Doncevic and Eike Cramer for fruitful discussions.
{Finally, the authors thank the anonymous reviewers and Tony Roberts for their valuable feedback and suggestions.}

\newpage
\singlespacing
\nolinenumbers

\end{document}